\documentclass[lineno]{jfm}

\usepackage{graphicx}
\usepackage{newtxtext}
\usepackage{newtxmath}
\usepackage{natbib}
\usepackage{tikz}
\usepackage{hyperref}
\usepackage{longtable}
\usepackage{pdflscape}
\usepackage{afterpage}
\usepackage{cleveref}
\hypersetup{
    colorlinks = true,
    urlcolor   = blue,
    citecolor  = blue,
}

\newcommand{\RomanNumeralCaps}[1]

\newcommand{\AxisRotator}[1][rotate=0]{%
    \tikz [x=0.25cm,y=0.7cm,line width=.1ex,-stealth,#1] \draw (0,0) arc (-150:150:1 and 1);%
}
\newcommand{\wT}{\overline{\langle wT \rangle} }
\newcommand{\T}{\overline{\langle T \rangle} }
\newcommand{\epsnu}{\overline{\langle \epsilon_\nu \rangle} }
\newcommand{\epsth}{\overline{\langle \epsilon_\theta \rangle} }

\definecolor{colorbar15}{rgb}{0.636364,0.000000,0.000000}
\definecolor{mygrey}{rgb}{0.6,0.6,0.6}

\title{
\nolinenumbers
Rotationally-affected Internally Heated Convection}

\author{
\nolinenumbers
Rodolfo Ostilla-M\'onico\aff{1}
  \corresp{\email{rodolfo.ostilla@uca.es}}
 \and Ali Arslan\aff{2}}

\affiliation{
\nolinenumbers
\aff{1}Dpto.~Ing. Mec\'anica y Dise\~no Industrial, Escuela Superior de Ingenier\'ia, Universidad de C\'adiz, Av.~de la Universidad de C\'adiz 10, 11519 Puerto Real, Espa\~na
\aff{2}Institute of Geophysics, ETH Zurich, 8092 Zurich, Switzerland}

\begin{document}
\maketitle

\begin{abstract}
\nolinenumbers
We study convection in a volumetrically heated fluid which is cooled from both plates and is under rotation through the use of direct numerical simulations. The onset of convection matches  similar systems and predictions from asymptotic analysis. At low rotation rates, the fluid becomes more organised, enhancing heat transport and increasing boundary layer asymmetry, whereas high rotation rates suppresses convection. Velocity and temperature statistics reveal that the top unstably stratified boundary layer exhibits behaviour consistent with other rotating convective systems, while the bottom boundary shows a unique interaction between unstable stratification and Ekman boundary layers. Additional flow statistics such as energy dissipation are analysed to rationalise the flow behaviour.  
\end{abstract}


\nolinenumbers
\section{Introduction}
\label{sec:intro}

Convection plays a decisive role in transport for geophysical and astrophysical environments. As such, a fundamental understanding of convection, as affected by rotation, will improve our knowledge of planetary and stellar dynamics. However, in comparison to boundary heating, convection driven by internal sources remains less studied. In planetary interiors, internal heating primarily due to the radioactive decay of isotopes is crucial for plate tectonics and dynamo generation \citep{schubert2001mantle,ricard2015}. In the cores of planets where dynamos typically occur, prior to inner core nucleation, internal heating by secular cooling drives the dynamo \citep{dormy2025rapidly,ROBERTS201557}. Crucially, for planetary and astrophysical dynamos, the convection that creates large-scale magnetic fields is constrained by rotation \citep{TILGNER2015183}, so an insight into internal heating with rotation is required.

Before considering rotation, the dynamics of internally heated convection (IHC) on its own warrants an introduction.
In contrast to Rayleigh-B\'enard convection (RBC), where fluid in a periodic plane layer is heated from below and cooled from above, with internal heating, the choice of heating and boundary conditions results in different flows \citep{Goluskin2016book}. For example, if net internal heating is zero and the boundaries are thermally insulating, a purely internally heated flow is generated devoid of the effects of boundary layers \citep{Bouillaut2019,Miquel2019,bouillaut2022}. This setup allows for diffusivity-free heat transport \citep{Lepot2018}. Whereas if the net heating is unity, two distinct flows occur based on the choice of boundary condition at the bottom. If the top is isothermal and the bottom an insulator (delimitated as IH3 \citep{Goluskin2016book}), all internal heating leaves via the upper boundary and the flow shares similarities with RBC. A Nusselt number, quantifying the ratio of the averaged total vertical heat transport to conduction, can be defined. However, if both the top and bottom are isothermal (IH1 in previous literature), heat can leave through both boundaries. In the latter scenario, convection quantifies the asymmetry in the heat flux out of the boundaries. As the mean conductive heat transport is zero, an exact Nusselt number does not exist, and the flow described contains notable differences to that of RBC.

A fundamental task in the study of fluid dynamics involves characterising the statistical properties of the flow as a function of the control parameters of the system \citep{Doering2019,lohse2023}. For IHC, this process is less thoroughly studied, especially for low Prandtl numbers, due to computational and experimental difficulties (see \cite{Goluskin2016book} for a review of experimental results).
The nondimensional control parameters are the Rayleigh number $R$, quantifying the ratio of destabilisation by internal heating to diffusion, the Prandtl number $Pr$, the ratio of viscous to thermal diffusivity and the aspect ratio $\Gamma$, comparing the horizontal to vertical scale of the system. 
Two properties of interest in IHC that quantify the effect of convective heat transport when the flow is laminar, chaotic and fully turbulent, are $\wT$ and $\overline{\langle T \rangle}$, the mean vertical heat transport by convection, and the mean temperature. For notation in this paper, $\langle \cdot \rangle$ denotes horizontal and time averaging,  and $\overline{\cdot}$ denotes vertical averaging. Unlike RBC, $\wT$ and $\T$ in IHC cannot be related \textit{a priori} and describe two separate emergent quantities of the system. Furthermore, the difference between IHC and RBC can be described by comparison of the horizontally averaged temperature and root mean square (RMS) velocity distributions. The IHC temperature profiles show the existence of a stably stratified boundary layer at the bottom with a thickness larger than the unstable thermal boundary layer at the top \citep{goluskin2012convection,goluskin2016penetrative}. 
The RMS velocity highlights the plumes from the upper boundary and asymmetric convective winds (relative to the midplane) within the domain.

The effect of rotation alters turbulent convection in several ways. Irrespective of buoyancy, the Coriolis force creates an asymmetry in velocity gradients and introduces an Ekman boundary layer \citep{greenspan1968,pedlosky}. The nondimensional Ekman number $E$, the ratio of viscous diffusion to rotation, gauges the effect of rotation and the creation of anisotropy in the flow. In the limit of zero Ekman number, the flow is subject to the Taylor-Proudman constraint, where the velocity is constant parallel to the axis of rotation and the dynamics are two-dimensional \citep{gallet2015}.

Rotating convection has only been studied in the plane layer when the heating source is the boundaries \citep{ecke2023}. The first studies on the onset of convection, summarised in \cite{chandrasekhar1961hydrodynamic}, highlighted the exact role of rotation in inhibiting the onset of convection and on the dominant length scales of the flow. Specifically, the critical Rayleigh number for linear instability scales as $E^{-4/3}$, while the dominant nondimensional length scales of the system go as $E^{1/3}$. These scalings were analytically demonstrated for internal heating in \cite{arslan2024internally}. Studies for rotating RBC indicate that the dominant lengthscale remains significant as the Rayleigh and Ekman numbers are in the turbulent regimes. Furthermore, numerical experiments have shown that the $E-R$ parameter space is rich with a variety of flow states (see \cite{ecke2023} and \cite{Kunnen04052021} for reviews). For sufficiently small $E$ and large enough $R$ for convection, the flow is geostrophically turbulent with characteristics different to buoyancy-driven turbulence \citep{Song2024}. A recent study of geostrophic turbulence with IHC is \cite{hadjerci2024}, with net zero heating in an entirely insulated domain.

It is worth highlighting that rotating IHC in alternative geometries is also relevant to geophysics. Consider a fluid in a sphere where the only possible buoyancy comes from within the domain, then linear stability analysis and simulations at onset and when supercritical have documented the flow states \citep{jones2000}. An important discovery has been that in spherical geometries, subcritical convection occurs for low Prandtl numbers ($Pr \leq 0.1$)\citep{kaplan2017,guervilly2016}. In addition to the sphere, studies of rotating spherical shells combine internal and boundary heating to drive convection \citep{gastine2016scaling}. Both geometries model the cores of the Earth at different periods in its history, albeit without the effects of magnetism. Returning to the plane layer, the energy method also predicts subcritical convection for non-rotating IHC \citep{Goluskin2016book}. However, the effect of rotation on nonlinear stability is unknown as rotating IHC with isothermal boundaries remains largely unexplored.

This work is motivated by the lack of studies of rotating IHC in the plane layer and, in part, by the analytical results obtained in \cite{arslan2024internally}. In \cite{arslan2024internally}, the author established the onset of convection for large $Pr$ ($Pr > 1$) and rotation (small $E$) and proved lower bounds on the heat flux out of the bottom boundary $\mathcal{F}_b$ and the mean temperature $\T$ in different areas of the $E-R$ space. That work uses the auxiliary functional method \citep{goulart2012global,Chernyshenko2022} to construct a variational problem for $\mathcal{F}_B$ and $\T$, the solution of which gives a lower bound in terms of $R$ and $E$. The method can prove sharp bounds on the long-time behaviour of ODEs and PDEs \citep{Fantuzzi2016b,Tobasco2018,Rosa2020}. In the case of the Navier-Stokes equations, quadratic auxiliary functionals have been used to prove bounds on turbulent flows (see \cite{Fantuzzi2022} for a detailed review). Although the method has proven sharp bounds for specific flows, in other cases, the bounds appear not to capture the dynamics. 

Bounds on $\wT$, for IHC, are an example where the sharpness of bounds remain unknown, even though the proofs require delicate mathematical constructions and estimates \citep{Arslan2021a,Arslan2021,kumar2021ihc,arslan2025a}. The bounds on long-time averages indicate that IHC has distinct features from RBC. Recent work shows that even when the internal heating is non-uniform, open questions remain, and that the limit of heating at the boundary remains unavailable \citep{Arslan2024b}. One means of testing the sharpness of bounds is via direct numerical simulations of a representative range of parameter values.
In general, proving results on rotating convection is challenging with a standard application of bounding methods because the Coriolis force does not appear in energy identities. The bounds proven for RBC or IHC, either consider the infinite Prandtl limit \citep{constantin1999r,Doering2001,Yan2004,arslan2024internally} or a reduced set of equations where time and length are rescaled by factors of $E^{1/3}$ \citep{grooms2014,Pachev2020}. 
While we cannot span the entire range of parameters, this work provides the first results for the variations of $\wT$ and $\T$ with $R$ and $E$ by direct numerical simulation in the rotationally-affected regime.

The paper is organised as follows. $\S$\ref{sec:nummeth} describes the simulation method. In $\S$\ref{sec:onsetconv}, we analyse the onset of convection in RIHC, while $\S$\ref{sec:flowmor} describes the resulting flow morphology. Global quantities are analysed in $\S$\ref{sec:globq}, while their relationship to upper bounds is discussed in $\S$\ref{sec:bounds}. The remaining results sections discuss the behaviour of local temperature and velocity statistics ($\S$\ref{sec:localstats}), boundary layer behaviour ($\S$\ref{sec:bls}) and dissipation rates ($\S$\ref{sec:diss}). Finally, $\S$\ref{sec:conc} briefly summarises the results obtained and appendix $\ref{sec:results}$ summarises the simulation data.

\section{Numerical Method}
\label{sec:nummeth}
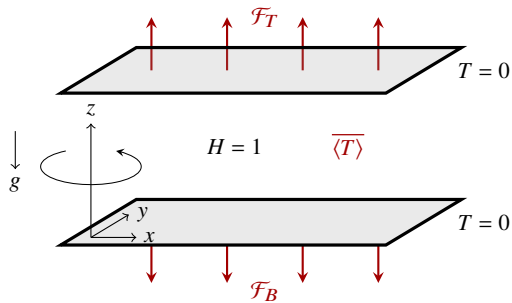
\begin{figure}
\centering
\begin{tikzpicture}[every node/.style={scale=0.95}]

    \draw [-stealth,colorbar15, thick] (-0.8,0) -- (-0.8,-0.5);
    \draw [-stealth,colorbar15, thick]  (0.2,0) -- (0.2,-0.5);
    \draw [-stealth,colorbar15, thick] (1.2,0) -- (1.2,-0.5);
    \draw [-stealth,colorbar15, thick]  (2.2,0) -- (2.2,-0.5);
    \draw [-stealth,colorbar15, thick]  (-0.8,2.3) -- (-0.8,3);
    \draw [-stealth,colorbar15, thick]  (0.2,2.3) -- (0.2,3);
    \draw [-stealth,colorbar15, thick]  (1.2,2.3) -- (1.2,3);
    \draw [-stealth,colorbar15, thick]  (2.2,2.3) -- (2.2,3);
    \draw[black,very thick, fill=mygrey, fill opacity = 0.2] (-2,2) -- (-1,2.6) -- (3.3,2.6) -- (2.3,2) -- cycle;
    \draw[black,very thick, fill=mygrey, fill opacity = 0.2] (-2,0) -- (-1,0.6) -- (3.3,0.6) -- (2.3,0) -- cycle;
    \draw[->] (-2.6,1.5) -- (-2.6,1) node [anchor=north] {$g$};
    \draw[->] (-1.6,0.1) -- (-1,0.1) node [anchor=west] {$x$};
    \draw[->] (-1.6,0.1) -- (-1.1,0.4) node [anchor=west] {$y$};
    \draw[->] (-1.6,0.1) -- (-1.6,1.6) node [anchor=south] {$z$};
    \node at (-1.6,1) {\AxisRotator[rotate=-90]};
    \node at (3.6,0.3) {$ T = 0 $};
    \node at (3.6,2.3) {$  T = 0 $};
    \node at (0.7,-0.6) {${\color{colorbar15}\mathcal{F}_B }  $};
    \node at (1.8,1.3) {${\color{colorbar15}\overline{\langle T\rangle}}  $};
    \node at (0.7,3) {${\color{colorbar15}\mathcal{F}_T }  $}; 
    \node at (0.3,1.3) {$ H = 1$};
    \end{tikzpicture}
\caption{A non-dimensional schematic diagram for rotating uniform internally heated convection. The upper and lower plates are at the same temperature, and the domain is periodic in the $x$ and $y$ directions and rotates about the $z$ axis. $\mathcal{F}_B$ and $\mathcal{F}_T$ are the mean heat fluxes out the bottom and top plates, $\overline{\langle T \rangle}$ the mean temperature, and $g$ is the acceleration due to gravity.}
\label{fig:schema}
\end{figure}

We consider a fluid layer confined by two horizontal walls separated a distance $d$, and periodic in both horizontal directions with equal periodicity lengths $\Gamma d$. The fluid has kinematic viscosity $\nu$, thermal diffusivity $\kappa$, density $\rho$, specific heat capacity $c_p$ and thermal expansion coefficient $\alpha$.  Gravity acts downwards with strength $g$, while the fluid rotates at rate $\Omega$ about the vertical axis and is uniformly heated internally at a volumetric rate $H$. Density variations are modeled with the Boussinesq approximation where they only contribute to buoyancy forces. Figure \ref{fig:schema} shows a schematic of the problem. 

To non-dimensionalise the system, we use $d$ as the characteristic length scale, $d^2/\kappa$ as the time scale and $d^2H/(\kappa \rho c_p)$ as the temperature scale \citep{roberts1967convection}. The velocity of the fluid $\textbf{u}(\textbf{x},t)=u(\textbf{x},t)\textbf{e}_1+v(\textbf{x},t)\textbf{e}_2+w(\textbf{x},t)\textbf{e}_3$ and temperature $T(\textbf{x},t)$ then satisfy the following non-dimensional equations:

\begin{gather}
    \nabla \cdot\textbf{u}=0, \label{eq:navierstokes1}\\
    \partial_t \textbf{u} + \textbf{u}\cdot\nabla\textbf{u} + E^{-1} \textbf{e}_3 \times\textbf{u} = -\nabla p + Pr \nabla^2\textbf{u} + Pr R T\textbf{e}_3, \label{eq:navierstokes2}\\
    \partial_tT + \textbf{u}\cdot\nabla T = \nabla^2 T +1 ,\label{eq:navierstokes3}
\end{gather}

\noindent subject to no-slip ($\textbf{u}=0$) and isothermal ($T=0$) boundary conditions at both walls. 

The system's response is governed by three non-dimensional parameters: the Rayleigh, Ekman and Prandtl numbers, defined as

\begin{equation}
    R = \frac{g\alpha Hd^5}{\rho c_p \nu \kappa^2}, \qquad E=\frac{\nu}{2\Omega d}, \qquad Pr=\frac{\nu}{\kappa}.
\end{equation}

\noindent The convective Rossby number, a parameter often used to directly compare the relative importance of buoyancy and rotation is obtained as $Ro = E\sqrt{R/Pr}$. 

This choice for non-dimensionalisation follows convention  \citep{goluskin2016penetrative, kazemi2022transition, arslan2024internally}, ensuring consistency with prior work, but it is not the only reasonable choice. Unlike Rayleigh-B\'enard, where the conventional non-dimensionalisation results in velocity and temperature scales are $\mathcal{O}(1)$ for $Pr=1$, internally heated convection exhibits velocity and temperature scales that vary with $R$. Our choice of non-dimensionalisation results in increasing velocity and decreasing temperature scales as $R$ increase, as discussed later. Alternative scalings, such as a free-fall-based velocity scale similar to RB would instead result in decreasing velocity and temperature scales, providing no inherent benefit while potentially introducing inconsistencies when comparing results to existing work. 

We solve equations \ref{eq:navierstokes1}-\ref{eq:navierstokes3} using the open-source, second-order centered finite difference AFiD \citep{van2015pencil}, which has been extensively validated for internally heated convection \citep{goluskin2016penetrative, kazemi2022transition}. The non-dimensional periodicity length $\Gamma$ is selected following guidelines by \cite{goluskin2016penetrative} to ensure that $\T$ and $\wT$ remain independent of domain size.

Following \cite{goluskin2016penetrative}, the resolution adequacy is measured by the exact relationships:

\begin{gather}
    \overline{\langle\epsilon_\nu\rangle}  \equiv \overline{\langle | \nabla^2\textbf{u}| \rangle} = R\overline{\langle wT\rangle}, \label{eq:exrel1}\\ 
    \overline{\langle\epsilon_\theta\rangle} \equiv \overline{\langle |\nabla^2T|\rangle} = \overline{\langle T \rangle} . \label{eq:exrel2}
\end{gather}

\noindent 
For the non-rotating cases, we find that the resolutions in \cite{goluskin2016penetrative} can be reduced without loss of accuracy. As rotation increases, the number of points in the $z$ direction must be increased to resolve the increasingly thinner Ekman boundary layers.

In this study, we will set $Pr=1$, and vary $(R,E)$ in the range $R\in[3.16\times 10^5,10^{10}]$ and $E\in[10^{-6},\infty)$. The full parameter space explored is discussed in the next section, with a summary of results and resolutions provided in the appendix. Except for the initial run, simulations are initialised from adjacent points in the ($R$,$E$) parameter space. Statistics are collected after transients to ensure temporal convergence of $\T$ and $\wT$ to less than 1$\%$. 

\section{Results}

\subsection{Onset of convection}
\label{sec:onsetconv}

We start our exploration by probing the onset of convection. As in Rayleigh-Bénard convection (RBC), rotation stabilises the system against vertical motion, delaying the onset of convection. However, unlike RBC, where a precise relationship between the Rayleigh and Ekman numbers determines the onset under rotation \citep{chandrasekhar1961hydrodynamic}, internally heated convection (IHC) permits only an asymptotic scaling due to the mathematical nature of the governing equations \citep{arslan2024internally}. This scaling follows $R_L\sim E^{-4/3}$ where $R_L$ is the critical Rayleigh number from linear instability analysis \citep{arslan2024internally}. The exponent $-4/3$ matches that of rotating RBC, reflecting a similarity between the two systems.

\begin{figure}
\centering
\includegraphics[width=.45\linewidth]{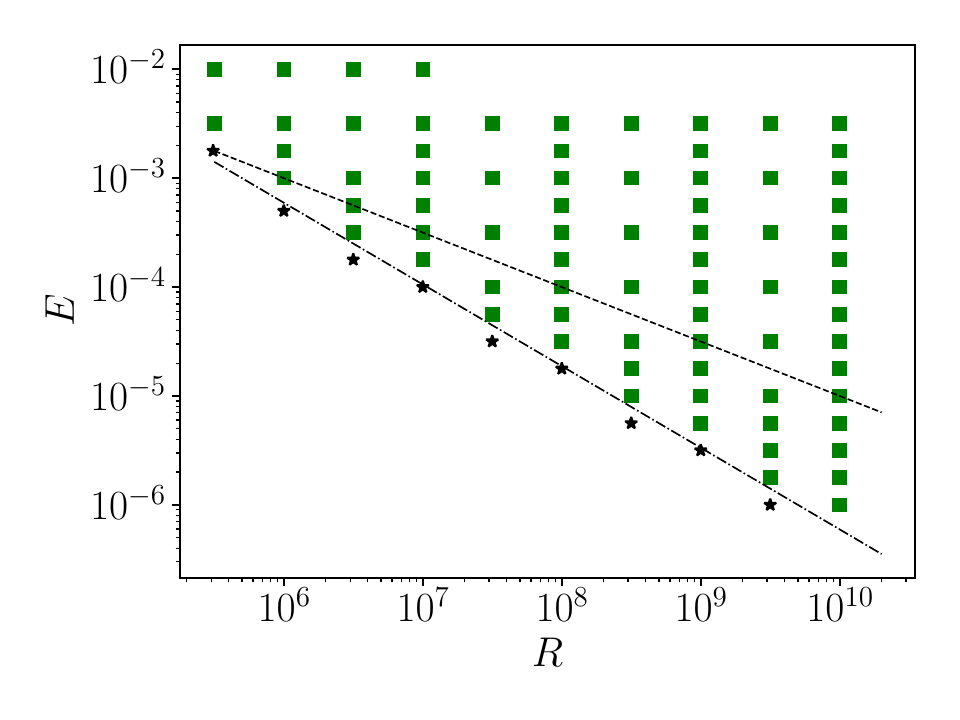}
\includegraphics[width=.45\linewidth]{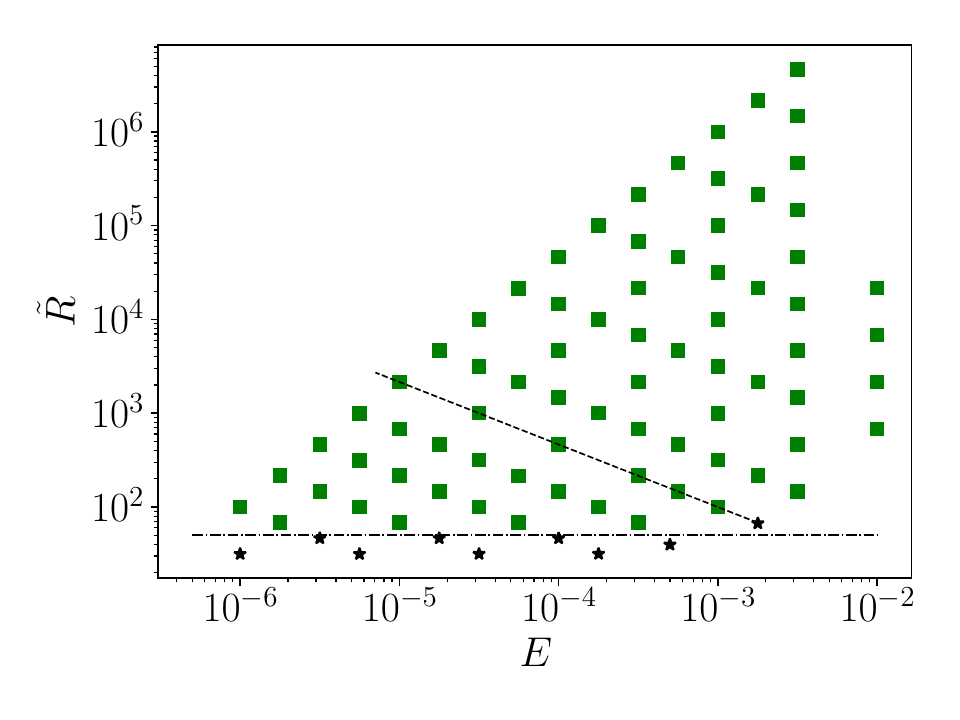}
\caption{Explored parameter space in ($R$,$E$) (left) and ($E$,$\tilde{R}$) (right) variables. Simulations conducted are marked as green squares, except when they produce purely conductive states and are marked as black stars. On both graphs, the dashed black line marks $Ro=1$ and the dash-dot line marks $\tilde{R}=50$. }
\label{fig:ek_ra_space}
\end{figure}

Using numerical simulations, we mapped the ($R$,$E$) parameter space to estimate the prefactor in this relationship. Figure \ref{fig:ek_ra_space} presents the simulations conducted, displayed both in ($R$,$E$) space and in ($E$,$\tilde{R}$) space, where $\tilde{R}=RE^{4/3}$ measures the degree of supercriticality in rotating IHC. The panels also show $Ro=1$, marking the region where rotation significantly influences the dynamics. We find that convection reliably sets in around $\tilde{R}\approx50$ when rotation becomes significant, independent of $R$.

When assessing system stability, sufficient simulation time is crucial for the flow to adjust to parameter changes. Increasing rotation initially suppresses turbulence, leading to a gradual rise in mean temperature, which destabilises the system. If given enough time (which may be rather long), turbulence may re-emerge. Thus, we define a purely conductive state as one where the mean temperature deviates by less than 1$\%$ from its conductive value.

\subsection{Flow Morphology}
\label{sec:flowmor}

\begin{figure}
\centering
\includegraphics[width=.3\linewidth]{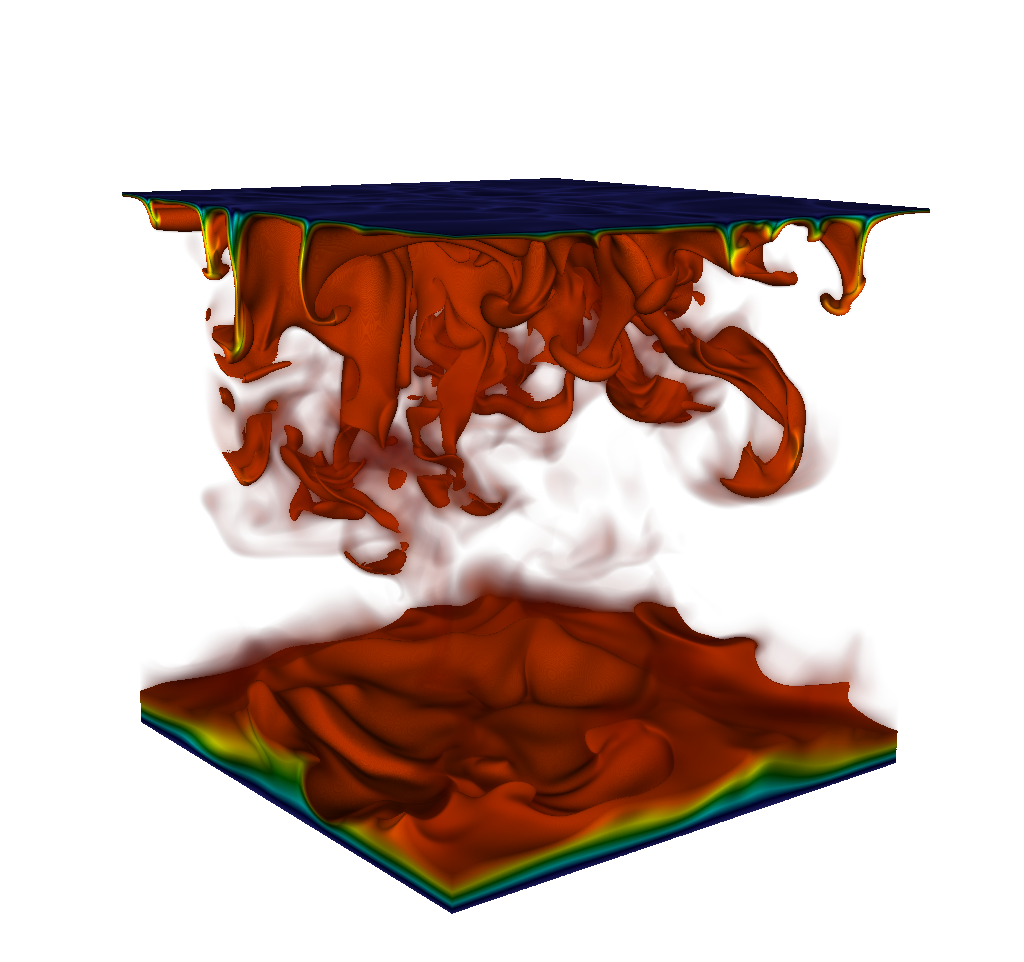}
\includegraphics[width=.3\linewidth]{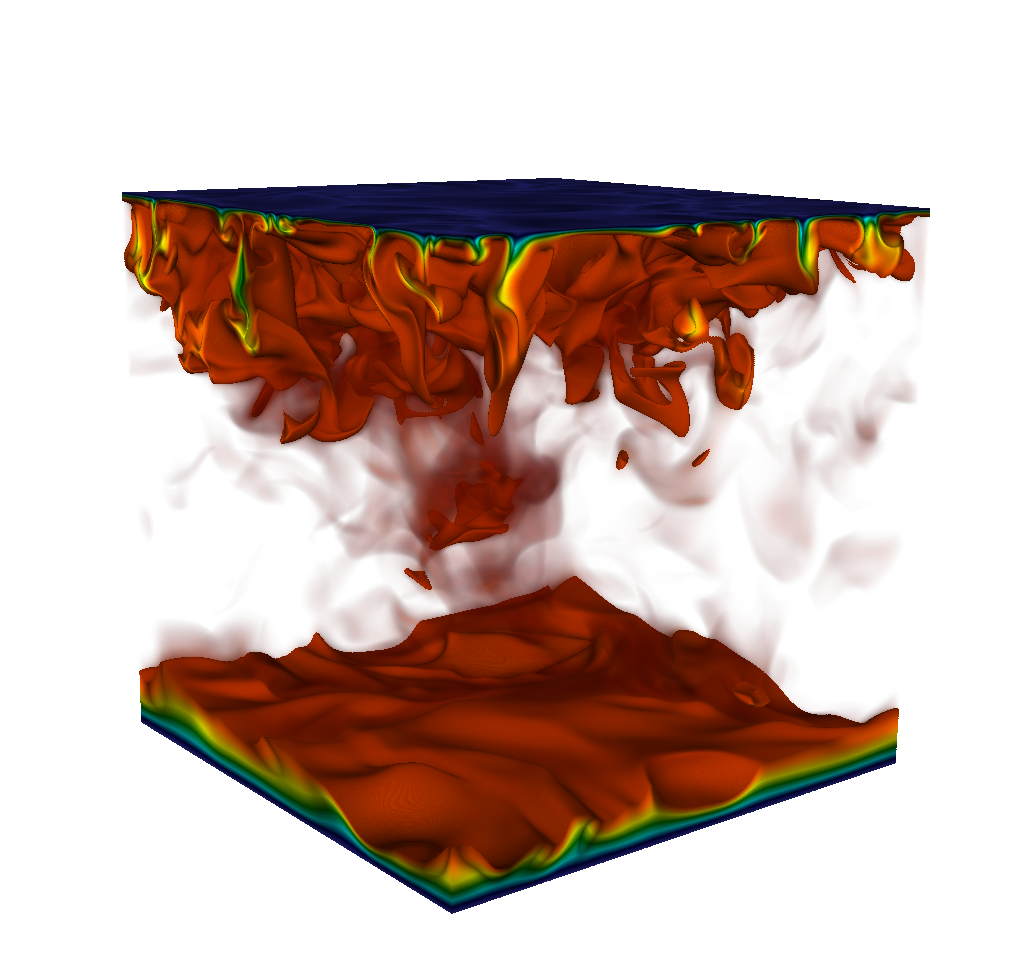}
\includegraphics[width=.3\linewidth]{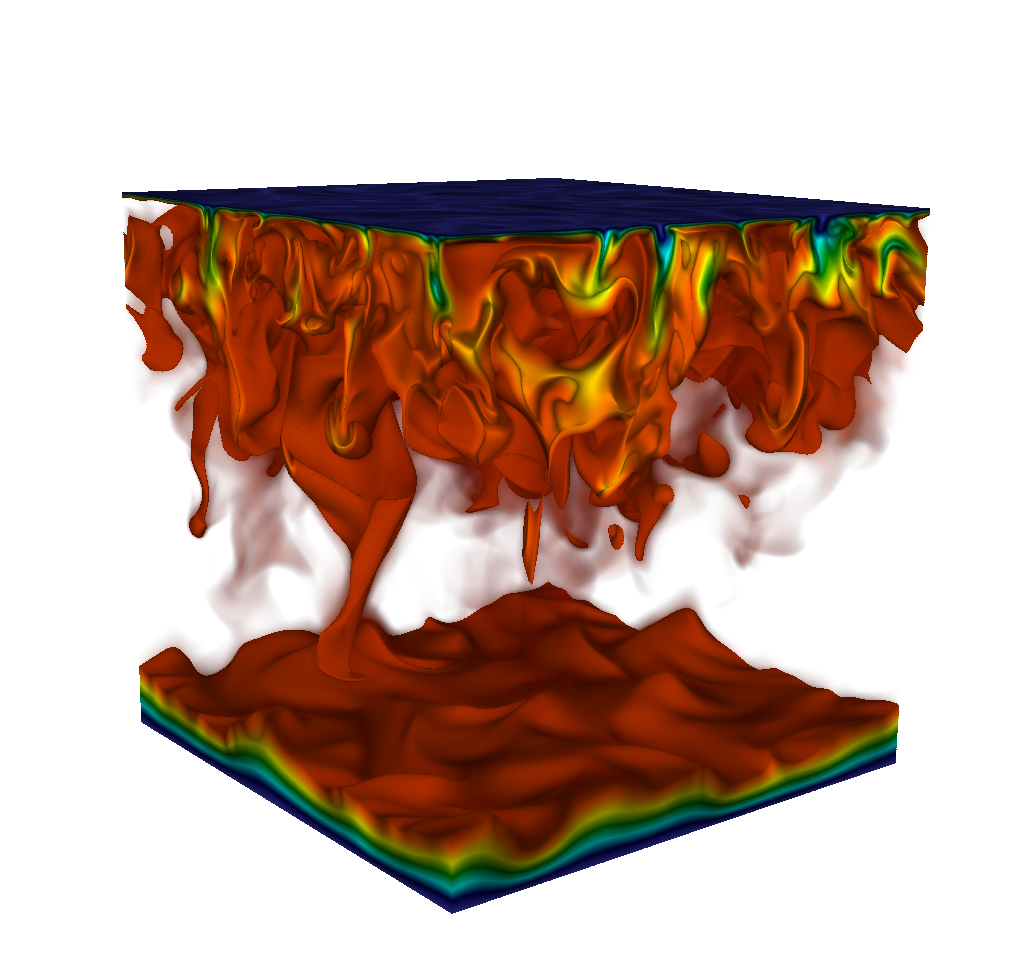}
\includegraphics[width=.3\linewidth]{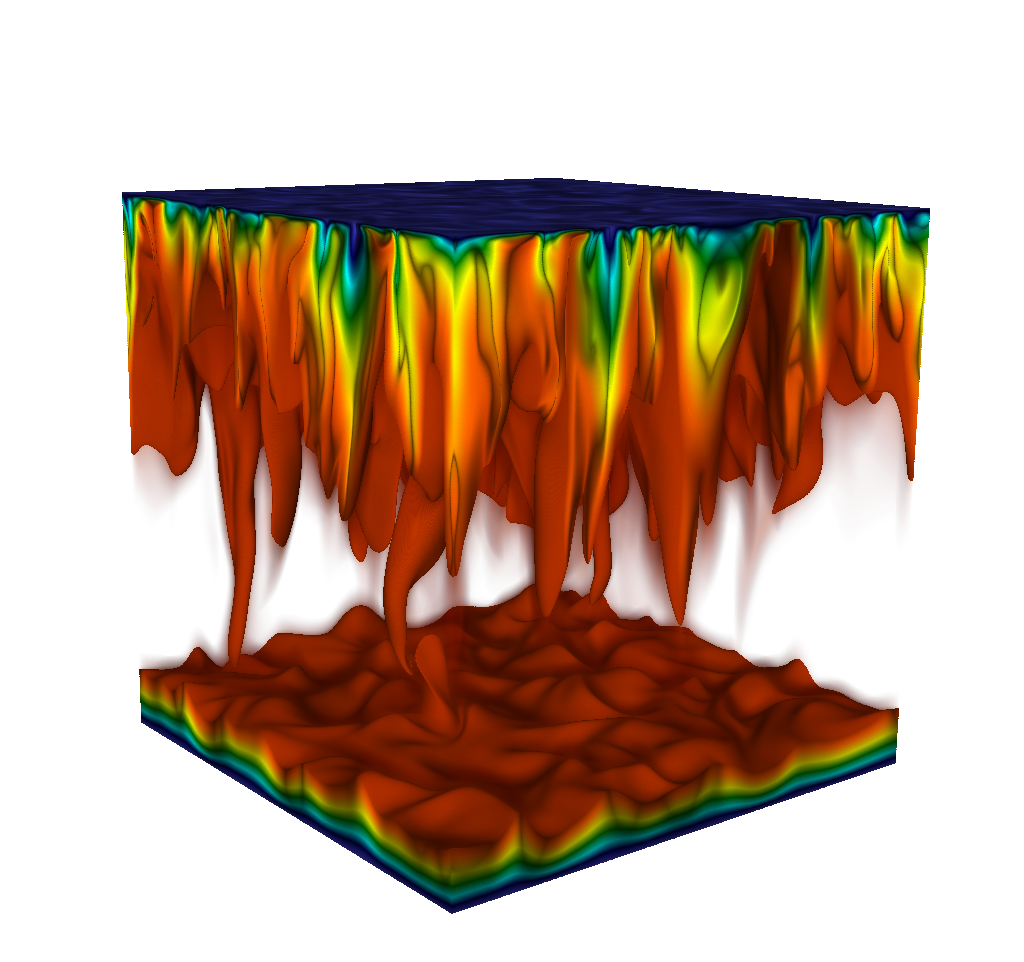}
\includegraphics[width=.3\linewidth]{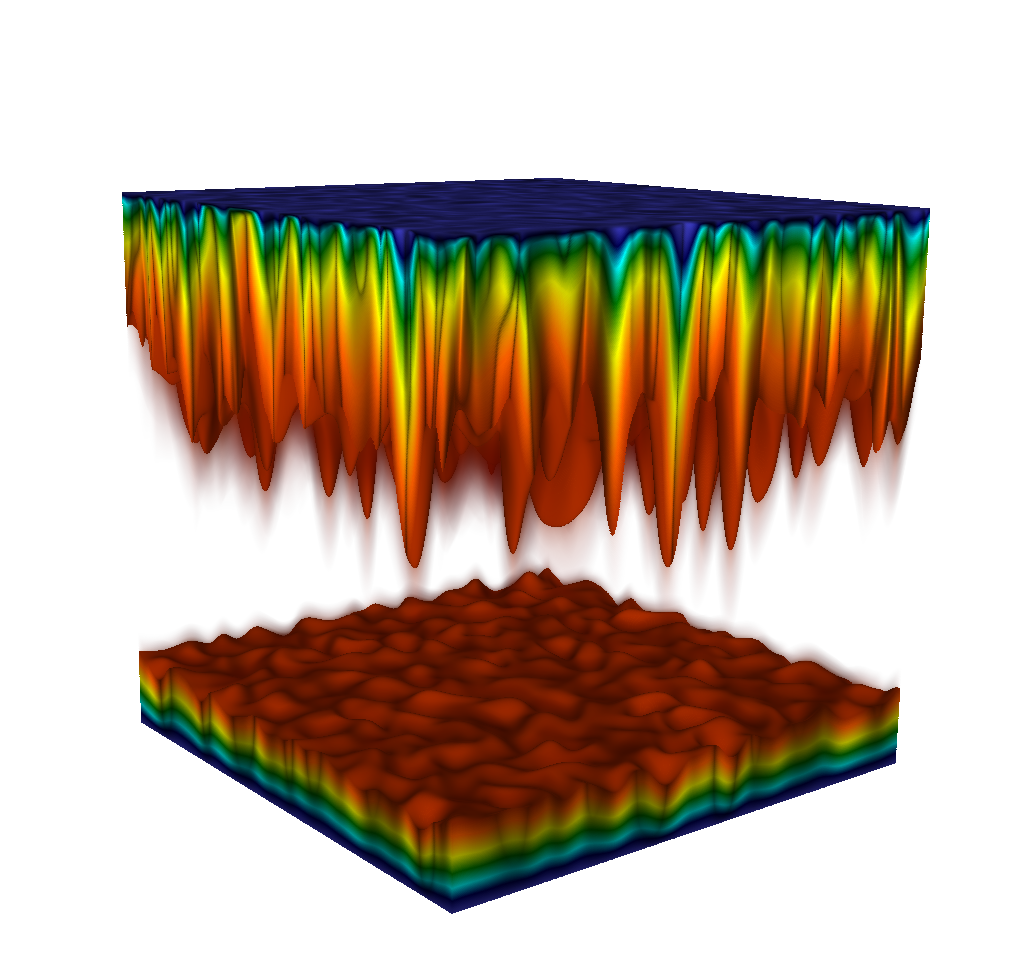}
\includegraphics[width=.3\linewidth]{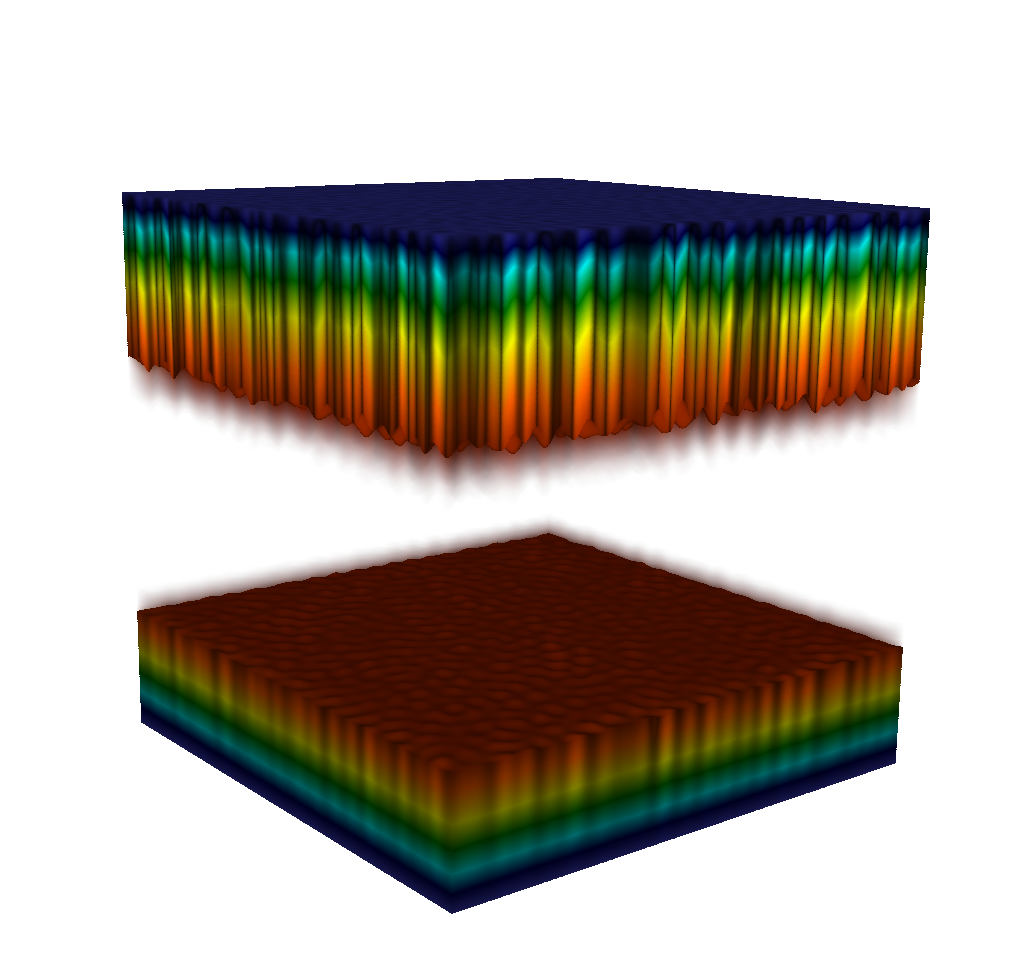}
\caption{Volumetric visualisation of the instantaneous temperature field for $R=10^{10}$ and from top-left to bottom-right: No rotation, $E=10^{-4}$ ($Ro=10$), $E=3.16\times10^{-4}$ ($Ro=3.16$), $E=10^{-5}$ ($Ro=1$), $E=3.16\times10^{-5}$ ($Ro=0.31$) and $E=10^{-6}$ ($Ro=0.1$). The flow can be seen to change structures from plume-dominated to column-dominated.}
\label{fig:flowvis}
\end{figure}

Figure \ref{fig:flowvis} illustrates how the flow evolves with increasing rotation (decreasing $E$) at $R=10^{10}$. Across all cases, plumes or other convective structures originate primarily in the top boundary layer, which is unstably stratified. This mixes the fluid, and produces turbulence which then mixes the stably stratified lower region. 

The flow structure changes significantly with rotation. Without rotation, convection is dominated by downwards plumes. As rotation increases, plumes elongate vertically and become more organised, eventually forming Taylor columns similar to those in rotating RBC. At $E=10^{-6}$, flow structures are highly constrained horizontally, and turbulence is nearly absent due to the low supercriticality ($\tilde{R}=100$). This means that our simulations are mainly in the parameter space corresponding to intermediate, or ``rotation-affected'' convection. While the smallest $E$ values considered correspond to $Ro=0.1$, the flow remains insufficiently supercritical to enter the geostrophic turbulence regime, instead exhibiting quasi-laminar structures.

\subsection{Global Quantities}
\label{sec:globq}

In this section we analyse the system response throughout the parameter space by examining three global parameters: $\mathcal{F}_B$, $\overline{\langle T \rangle}$ and the wind Reynolds number $Re_w=U_wd/\nu$, where $U_w$ is a characteristic velocity for the wind defined as:

\begin{equation}
 U_w = \overline{\langle u^2+v^2+w^2\rangle}^{1/2}.
\end{equation}

\begin{figure}
\centering
\includegraphics[width=.32\linewidth]{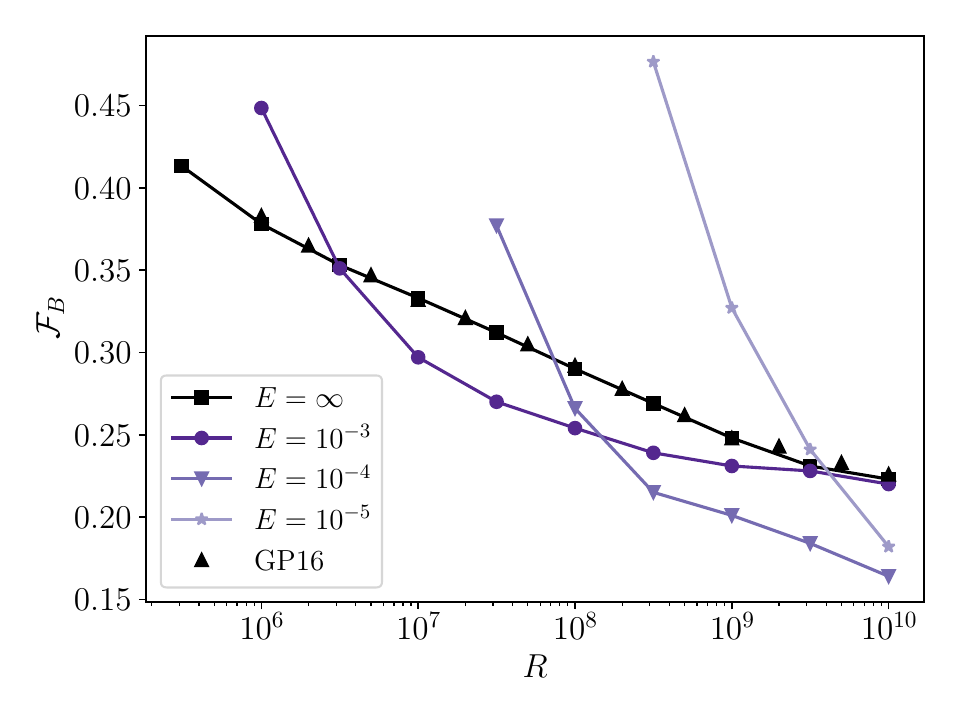}
\includegraphics[width=.32\linewidth]{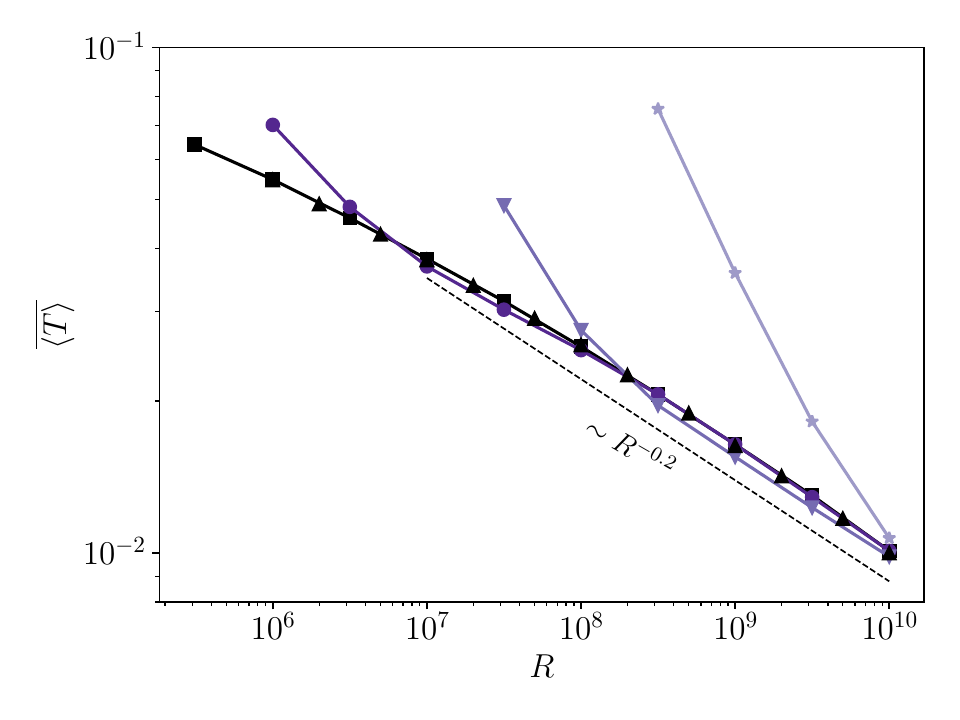}
\includegraphics[width=.32\linewidth]{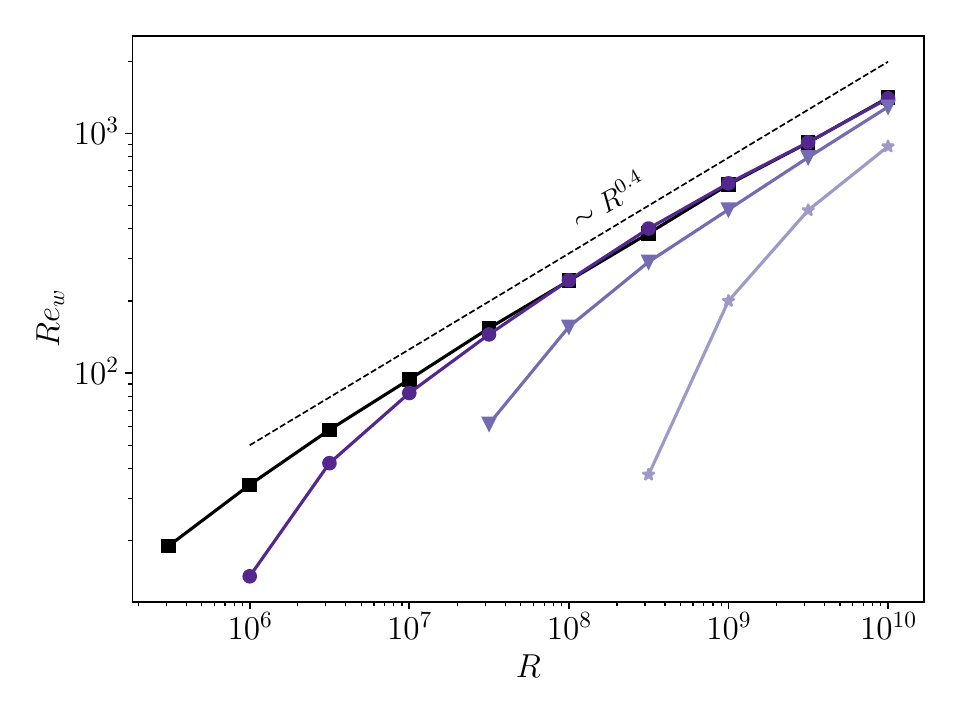}\\
\includegraphics[width=.32\linewidth]{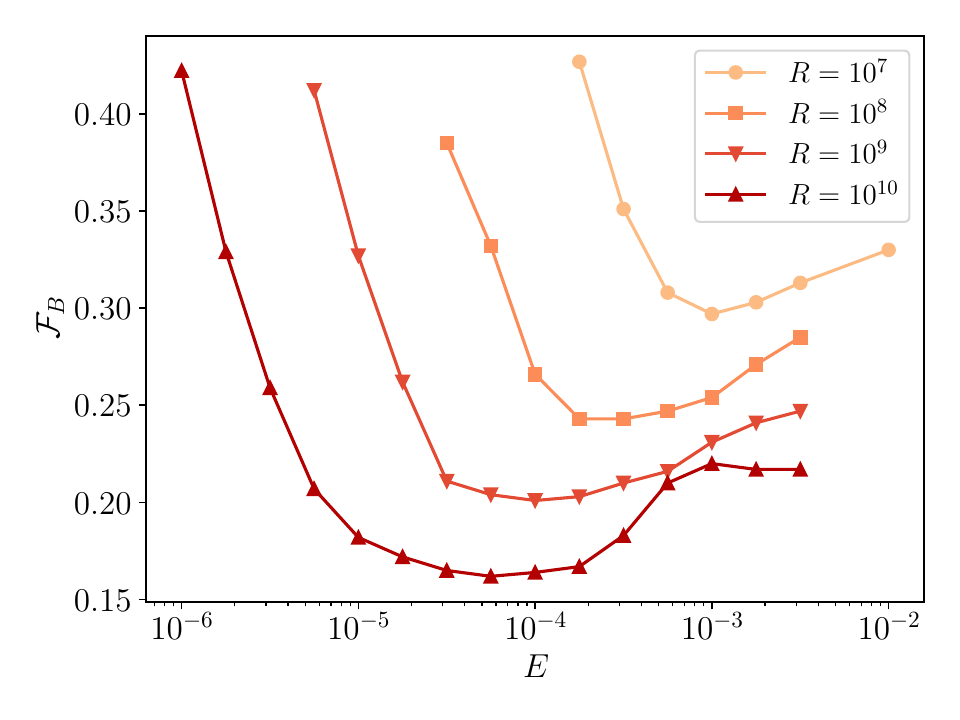}
\includegraphics[width=.32\linewidth]{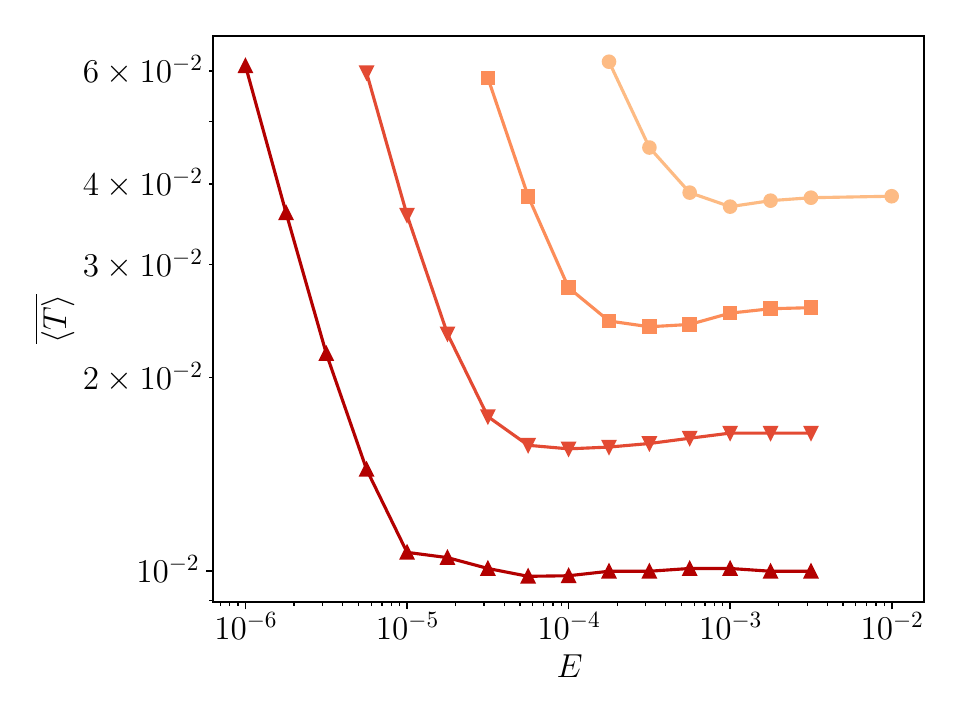}
\includegraphics[width=.32\linewidth]{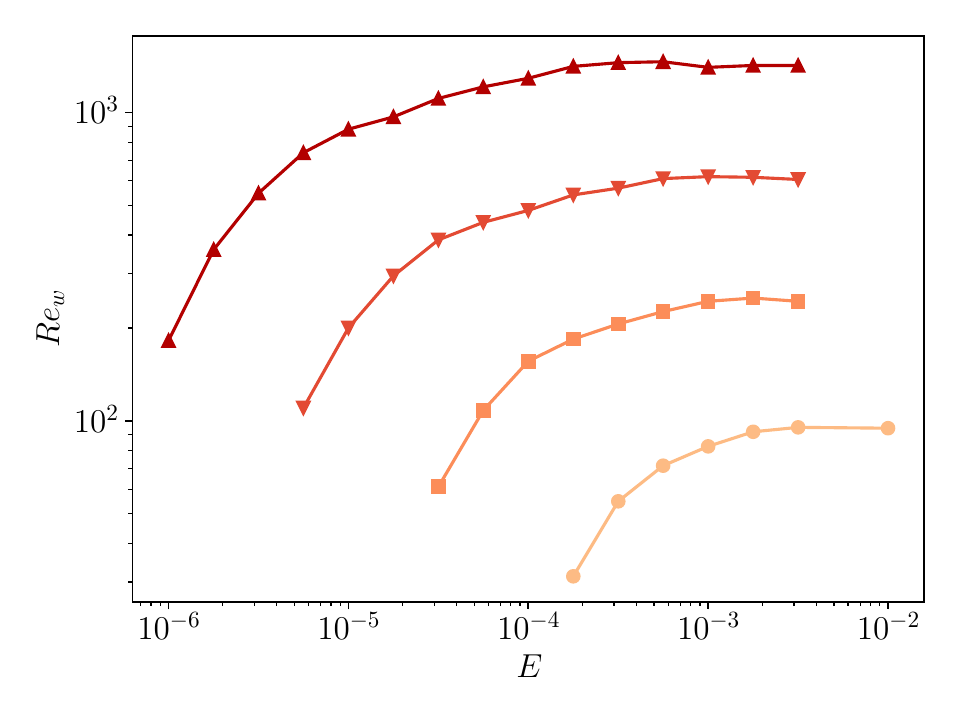}
\caption{Global responses. Top row: $\mathcal{F}_B$ (left), $\overline{\langle T \rangle}$,  (center) and $Re_w$ (right) against $R$ for several values of $E$. Bottom row: same quantities against $E$ for several values of $R$.}
\label{fig:globalqs}
\end{figure}

The left panels of figure \ref{fig:globalqs} shows the behaviour of $\mathcal{F}_B$, i.e.~the fraction of heat transported through the bottom plate, through the ($R$,$E$) parameter space. This is the quantity conventionally used when representing mathematical upper bounds. In our setup of IHC, all heat must leave through the boundaries. Therefore, the fractions corresponding to heat leaving from the top and bottom boundaries must be equal to unity.  Vertical convection, $\wT$, causes more heat to leave through the top boundary than the bottom. The exact relationships which link the variables are:

\begin{equation*}
 \mathcal{F}_B + \mathcal{F}_T = 1 , \qquad \mathcal{F}_B = \frac12 - \wT, \qquad \mathcal{F}_T=\frac{1}{2}+\wT\, .
\end{equation*}

\noindent Therefore, representing $\wT$ is equivalent to representing $\mathcal{F}_B$. 
 
Results for non-rotating IHC ($E=\infty$) from \cite{goluskin2016penetrative} are included for comparison, and an excellent agreement between both sets of simulations can be appreciated, despite the lower resolutions used in this manuscript. Two dependencies are apparent from the graph: first, that as $R$ increases, the fraction of heat transported through the bottom plate $\mathcal{F}_B$ decreases regardless of the value of $E$. While this decrease is slow, going from $40\%$ at the lowest $R$ to the lowest value seen in the graph of $15\%$ for rotating IHC, the monotonic trend of $\mathcal{F}_B$ is apparent. 

While it may seem intuitive that increased driving leads to higher convective vertical transport and more heat leaving from the top, this is not the case for 2D IHC \citep{goluskin2012convection}. \cite{goluskin2016penetrative} postulated that this was due to the existence of a competing effect: increased turbulence leads to shear-driven mixing of the bottom boundary layer, which allows for more heat to escape. In 2D simulations, the geometrical constraints on the flow result in organisation. A large-scale circulation develops, which heavily increases the shear-driven mixing. In 3D, this circulation does not form for the non-rotating case. In our simulations, we do not observe non-monotonic behaviour of $\mathcal{F}_B(R)$, despite rotation changing the flow morphology, further confirming that this non-monotonicity is a product of the 2D simulations. 

The second behaviour observed is that there are regions of parameter space where rotation results in lower values of $\mathcal{F}_B$. This is especially pronounced for the larger values of $R$ seen in the bottom left panel of Figure \ref{fig:globalqs}. This means that the overall level of vertical convective transport increases. Furthermore, the value of $E$ where $\mathcal{F}_B$ is minimum can be observed to depend on $R$. We will return to this phenomena, which is akin to increases in the heat transport seen for rotating RB at comparable values of the Rayleigh number.

The middle panels of figure \ref{fig:globalqs} show the behaviour of the volumetric averaged temperature $\overline{\langle T \rangle}$ in the parameter space. Both patterns of behaviour seen in $\mathcal{F}_B$ are also present here: the mean temperature decreases with $R$, and certain finite values of $E$ lead to decreases of the mean temperature when compared to those for the non-rotating case. We also note that the mean temperature scales roughly like $R^{-0.2}$. This scaling is only an approximation and disappears when the values are examined closely. However, what is clear is that the temperature scale of the system diminishes as $R$ increases, due to the higher level of turbulent convection.

Finally, the right panels of Figure \ref{fig:globalqs} show the behaviour of the wind Reynolds number. For this quantity, the dependence on driving parameters is simpler: $R$ increases velocity fluctuations, while decreasing $E$ damps these. No `optimal' rotation rate which increases $Re_w$ is observed, demonstrating that the variations of $\mathcal{F}_B$ and $\T$ are due to flow organisation and not to increased wind. We also note that $Re_w$ very approximately scales as $Re_w\sim R^{0.4}$. While the exact value of this exponent is not important, what is clear is that the scaling is less steep than one which would correspond to a free-fall velocity, i.e.~$R^{0.5}$. This is because the velocity scale is a response of the system, and because the temperature scale is decreasing with $R$ (as seen from the behaviour of $\T$), the forcing and hence the scaling exponent of $Re_w$ is reduced as a consequence. We note that our results are different from the scalings observed for $Re_w$ in 2D simulations by \cite{Wang2020}, which see an approximate free-fall scaling. This can again be attributed to the large differences in velocity statistics observed in IHC when comparing 2D and 3D simulations \citep{goluskin2016penetrative}. 

\begin{figure}
\centering
\includegraphics[width=.45\linewidth]{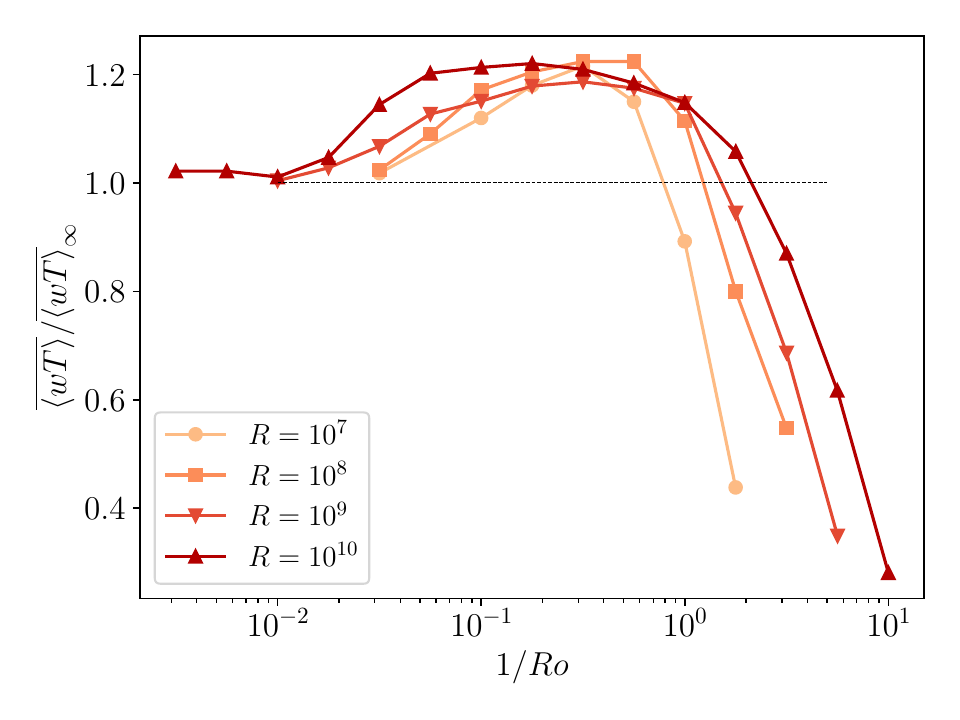}
\includegraphics[width=.45\linewidth]{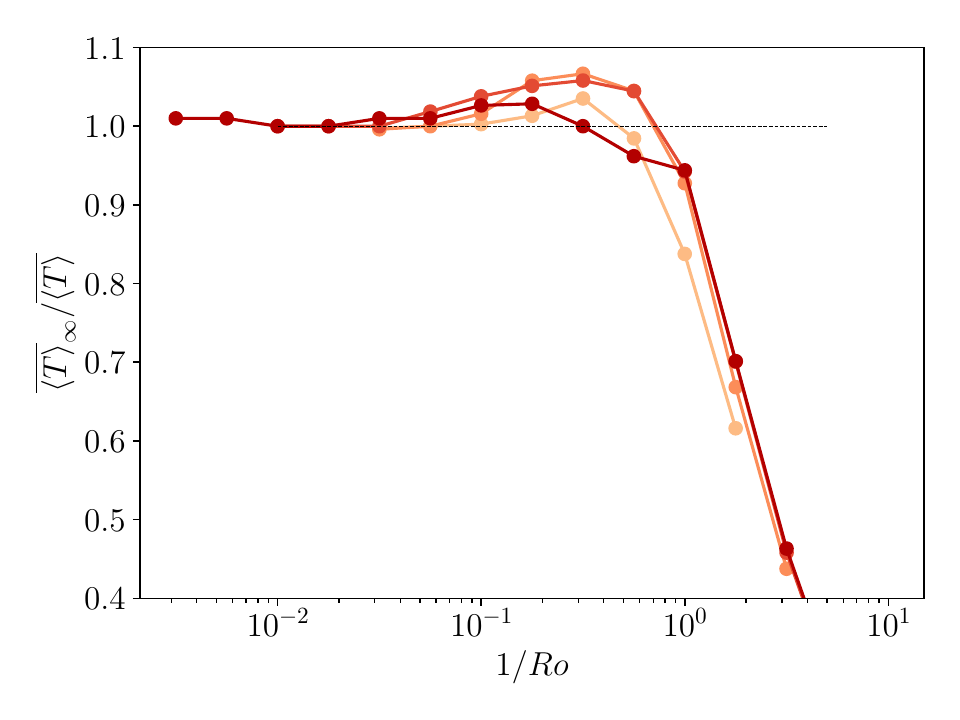}
\caption{Optimal transport: Normalised $\langle wT \rangle$ (left) and $\overline{\langle T \rangle}$  (right) against $1/Ro$ for several values of $R$.}
\label{fig:opttrans}
\end{figure}

To further analyse the phenomena of optimal transport, in Figure \ref{fig:opttrans} we show the behaviour of $\wT$ and $1/\T$ normalised by their values for the non-rotating case ($E=\infty$) as a function of the inverse Rossby number. This representation allows for a better collapse in the independent variable, while displaying the relative increases more clearly. Three regimes can be observed: (i) for small values of $1/Ro$, rotation does not affect the flow, (ii) for intermediate values of $1/Ro$, rotating the flow organises the motions and increases convective transport resulting in larger values of $\wT$ and smaller values of $\T$, (iii) for large values of $1/Ro$, rotation suppresses convection resulting in smaller $\wT$ and larger $\T$. We note that the collapses are not perfect: the changing velocity scale is reflected in slight shifts in the curves.

We also note that the relative increases in $\wT$ are much larger than those of $1/\T$, reaching up to $25\%$ in the former, while remaining below $10\%$ for the latter. Furthermore, the increases in $1/\T$ are very $R$ dependent, and are almost eliminated for the highest value of $R$. This is inline to what is seen for the Nusselt number in rotating RBC: a ``rotation-affected'' regime, where the Ekman boundary layers formed at the no-slip plates transfer heat effectively and cause small increases in the heat transfer that vanish at higher thermal drivings due to changes in flow organisation \citep{Kunnen04052021}. This change in flow organisation can be also appreciated in Figure \ref{fig:flowvis}. Furthermore, for $R=10^{10}$ in our simulations, the heat transfer increase has practically disappeared from the right panel of Figure \ref{fig:opttrans} in line with what is expected from rotating RBC. 

However, for $\wT$ the increases remain for all $R$. As mentioned earlier, the asymmetry between heat escaping the top and bottom boundary layers is also driven by the shear mixing of the stably stratified bottom. The convective structures present in rotating turbulence result in smaller $Re_w$, and mix the bottom less well. This second mechanism is probably behind the persistent increases in $\wT$: vertical convection compensates the decreased amount of heat escaping the bottom plate, and increases the asymmetry between boundary layers.

\subsection{Comparison to bounds}
\label{sec:bounds}

The global quantities can also be compared to the rigorous bounds proven with the methods mentioned in the introduction. We therefore include a brief comparison with the results proven in \cite{arslan2024internally}. An exact evaluation of the scaling laws was not made from the results, nevertheless, a qualitative comparison with the rigorous bounds are possible. Also, we keep in mind that the bounds in \cite{arslan2024internally} are proven for \cref{eq:navierstokes2} in the limit of infinite $Pr$, while our simulations are at $Pr=1$. However, in \cite{tilgner2022} the author shows that for fixed $R$ and $E$, the difference between the bounds in the limit of infinite and finite $Pr$, for RBC, contains the correction $\sim (\sqrt{1 + Pr^{-2}} +  Pr^{-1})^2$. Further still for $Pr=1$, there is an additional $R$ and $E$ dependence that changes the bounds. The equivalent correction for rotating IHC is unknown.

The range of parameter space explored in this work, as shown in \cref{fig:ek_ra_space}, falls within regimes III and IIIa of \cite{arslan2024internally}, see table 1 and figure 8. 
For $\mathcal{F}_B$, the bounds have been proven for regime III, the rotationally-affected regime, and are that $\mathcal{F}_B \gtrsim R^{-2/3}E^{2/3} + R^{-1/2}E^{1/2} \ln{|1 - R^{-1/3}E^{1/3}|}$. Comparing to the first row of \cref{fig:globalqs}, for fixed $E$, $\mathcal{F}_B$ decreases with $R$, as in the bound. For fixed $R$ the lower bound matches the behaviour only higher $E$ (slower rotation). On the other hand, the bounds for $\T$, in the region where we have carried out simulations, are split into regimes III where $\T \gtrsim R^{-2/7}E^{2/7}$ and IIIa where $\T \gtrsim R^{-1}E^{-1}$. Regime IIIa captures the increased influence of rotation on convection. The middle row of \cref{fig:globalqs} matches the behaviour of the bounds. In particular, fixing $R$ there are, at the very least, two distinct behaviours of $\T$ as $E$ varies, and this correlates with the bounds in III and IIIa. Therefore, the bounds on $\T$ qualitatively match the simulation results, while those for $\mathcal{F}_B$ can be improved for smaller $E$.

\subsection{Temperature and velocity statistics}
\label{sec:localstats}

\begin{figure}
\centering
\includegraphics[width=.45\linewidth]{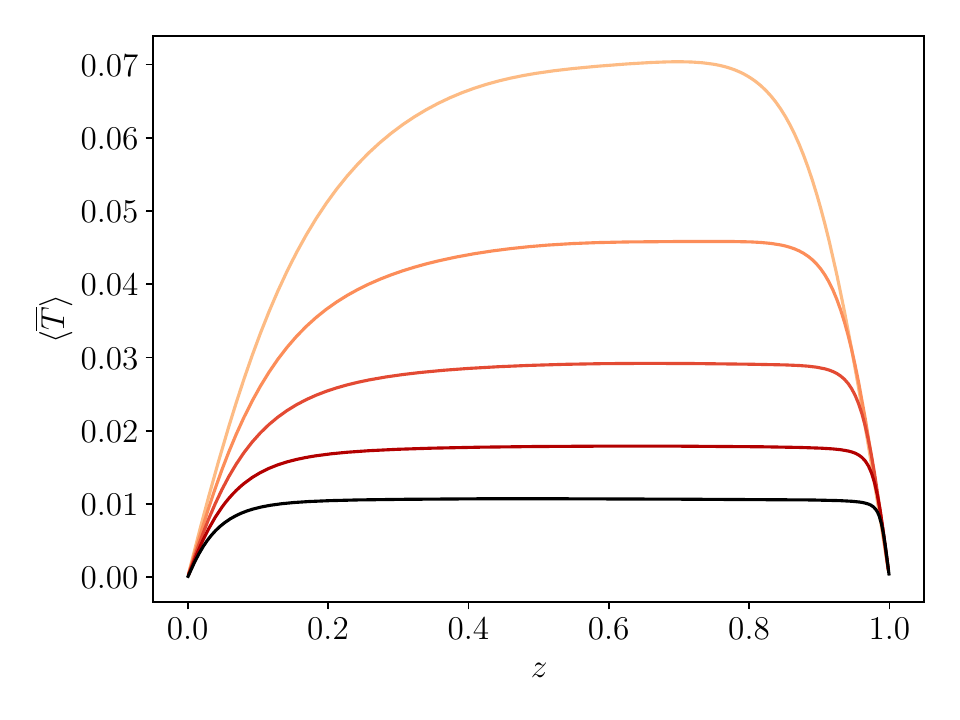}
\includegraphics[width=.45\linewidth]{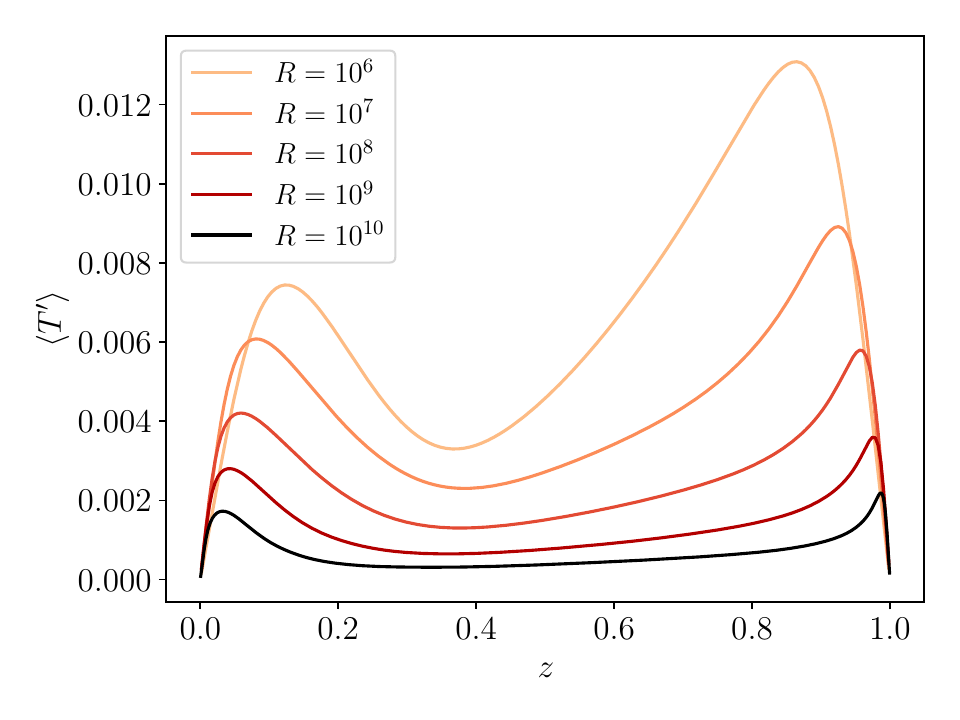}
\includegraphics[width=.45\linewidth]{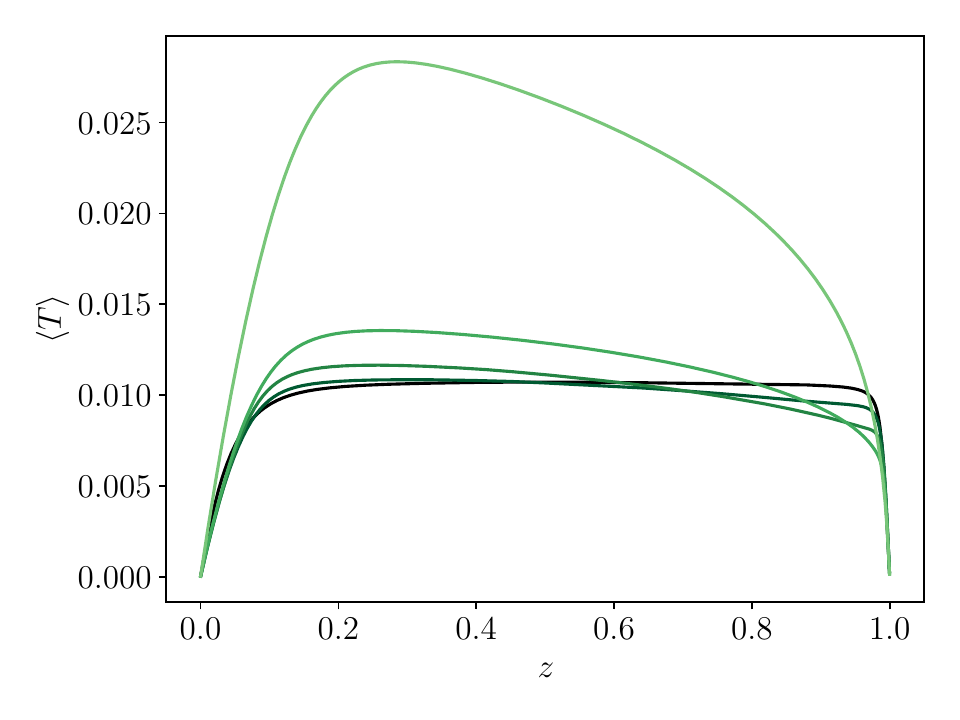}
\includegraphics[width=.45\linewidth]{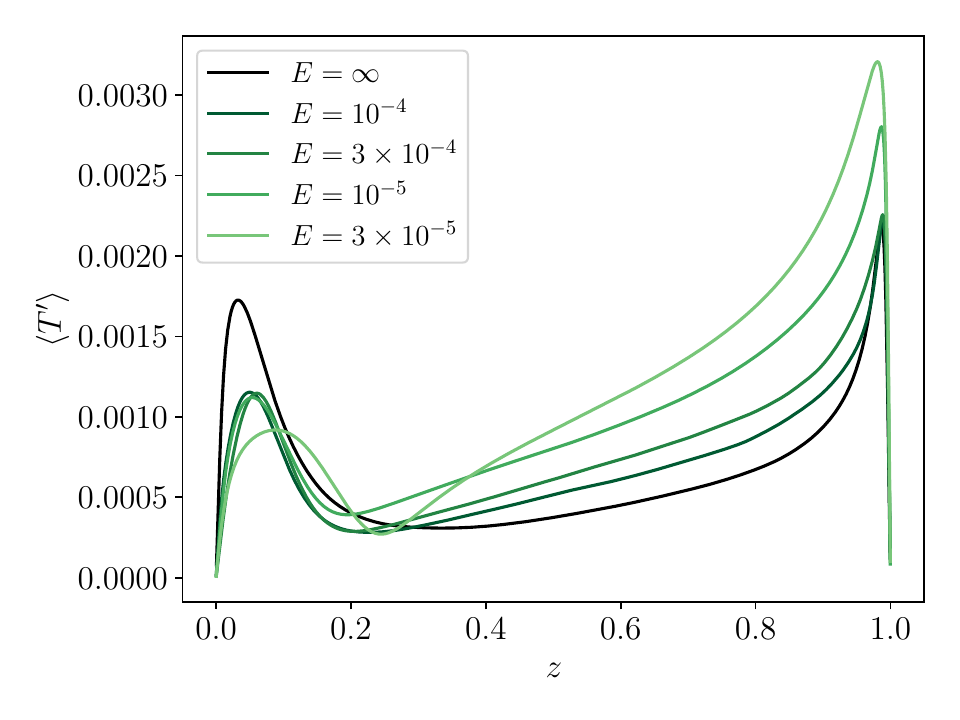}
\caption{Mean $\langle T \rangle$ (left column) and fluctuation $\langle T^\prime \rangle$ (right column) profiles for several values of $R$ in the non-rotating case (top row) and for several values of $E$ at $R=10^{10}$ (bottom row).}
\label{fig:tprofiles}
\end{figure}

To further understand the results obtained, we turn to an examination of the local behaviour of the fluid quantities. Starting with temperature, Figure \ref{fig:tprofiles} shows the mean $\langle T \rangle$ and RMS $\langle T^\prime \rangle = \sqrt{\langle T^2\rangle - \langle T\rangle^2 }$ temperature profiles for several values of $R$ and $E$. In the top panels, both mean temperature and fluctuations generally decrease with increasing $R$,  consistent with the decline in $\T$. Without rotation, the mean profiles exhibit a well-mixed bulk with a nearly constant temperature, a sharp gradient in the top boundary layer, and a more gradual gradient in the stably stratified bottom boundary layer. As $R$ increases, the well-mixed region expands, and the temperature profile flattens.

In contrast, the fluctuations show two peaks: a more pronounced one near the top boundary layer and a weaker one near the bottom. The minimum fluctuation region is located between $z=0.2$ and $z=0.4$, and shifts downward as $R$ increases. We identify this minimum with the interface between stably and unstably stratified layers. This is a better diagnosis than the position of the mean temperature maximum, as the latter is located closer to the top plate than the bottom plate due to the temperature gradient in the bulk, and does not correspond to the physical reality of the system.

\begin{figure}
\centering
\includegraphics[width=.45\linewidth]{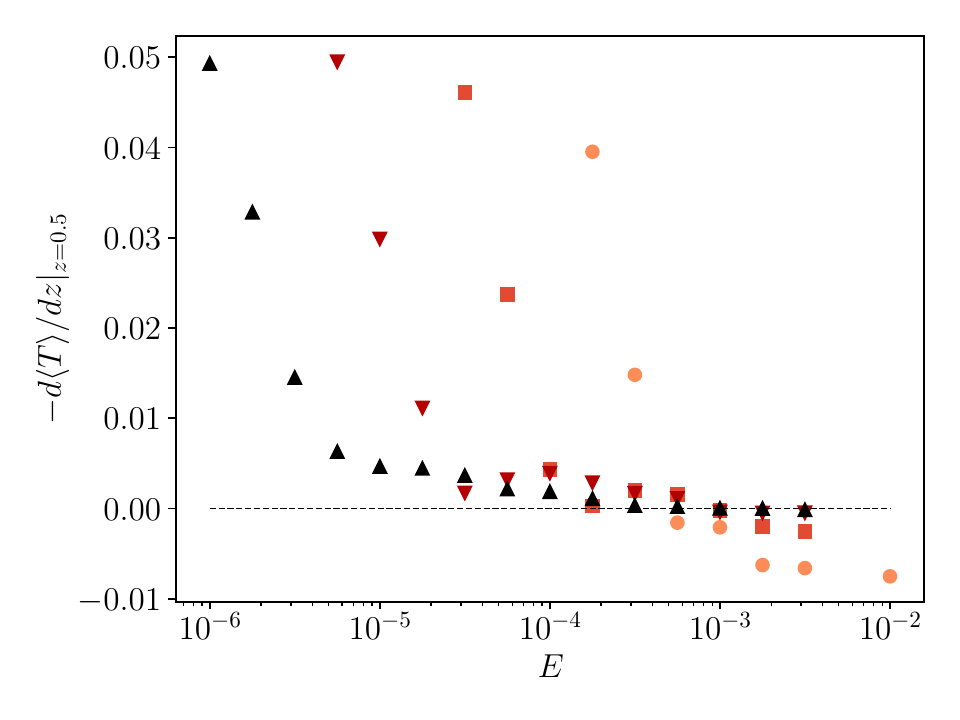}
\includegraphics[width=.45\linewidth]{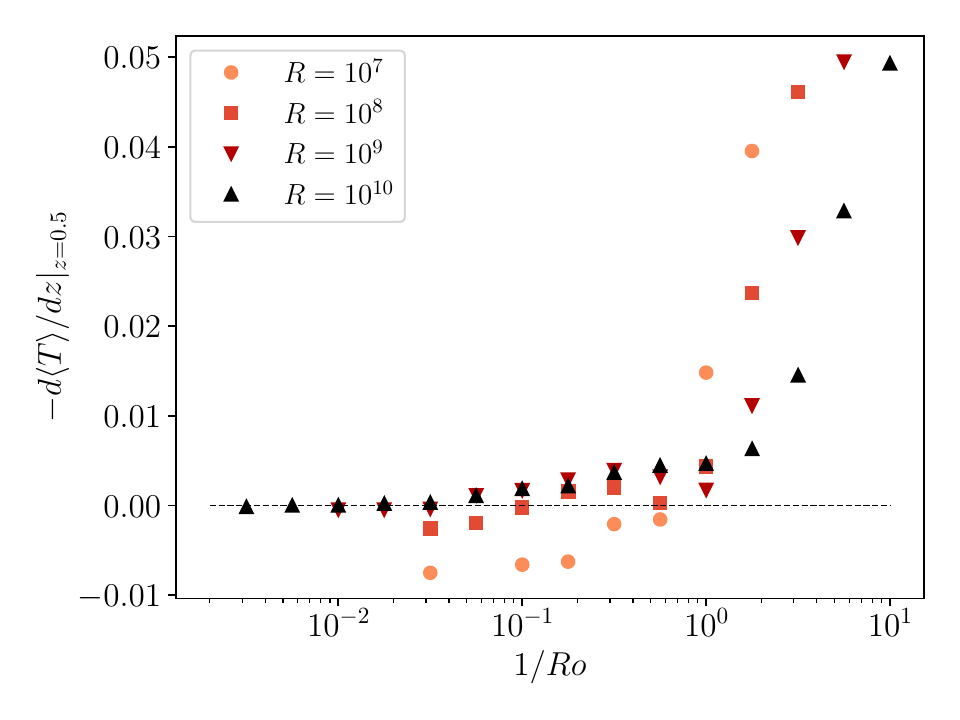}
\caption{Bottom panels: Temperature gradient in the mid-gap as a function of $E$ (left) and $1/Ro$ (right). }
\label{fig:tgrad}
\end{figure}

With rotation, the shape of the mean temperature profiles remain largely unchanged except for the emergence of a bulk temperature gradient, which becomes steeper as $E$ decreases. This gradient, opposite in sign to the weak positive gradient seen without rotation, resembles a phenomenon observed in rotating Rayleigh-Bénard convection \citep{Kunnen04052021} and is quantified in Figure \ref{fig:tgrad}. Initially, the bulk remains well mixed with a slight positive gradient at low $R$, but as rotation strengthens, the gradient reverses. Once $1/Ro>1$, the gradient grows rapidly in absolute terms.

Temperature fluctuations also respond to rotation: fluctuations in the bottom boundary layer decrease, while those in the top increase. This suggests that rotation reduces the mixing in the bottom region, and supports the idea mentioned earlier that increases in $\wT$ are driven by less thorough mixing of the stably stratified layer.

\begin{figure}
\centering
\includegraphics[width=.45\linewidth]{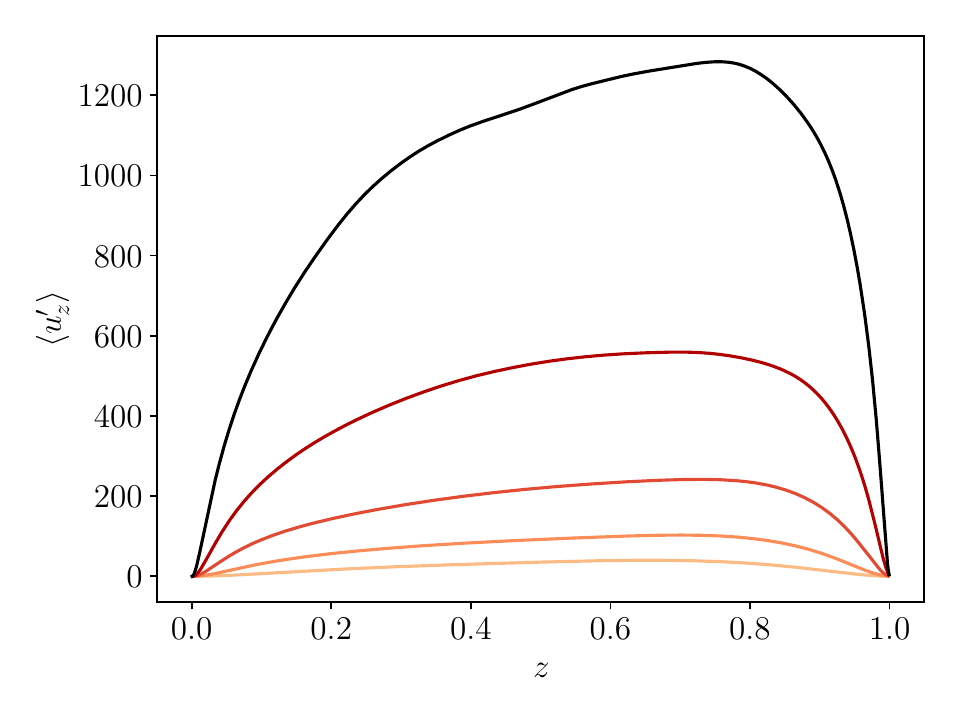}
\includegraphics[width=.45\linewidth]{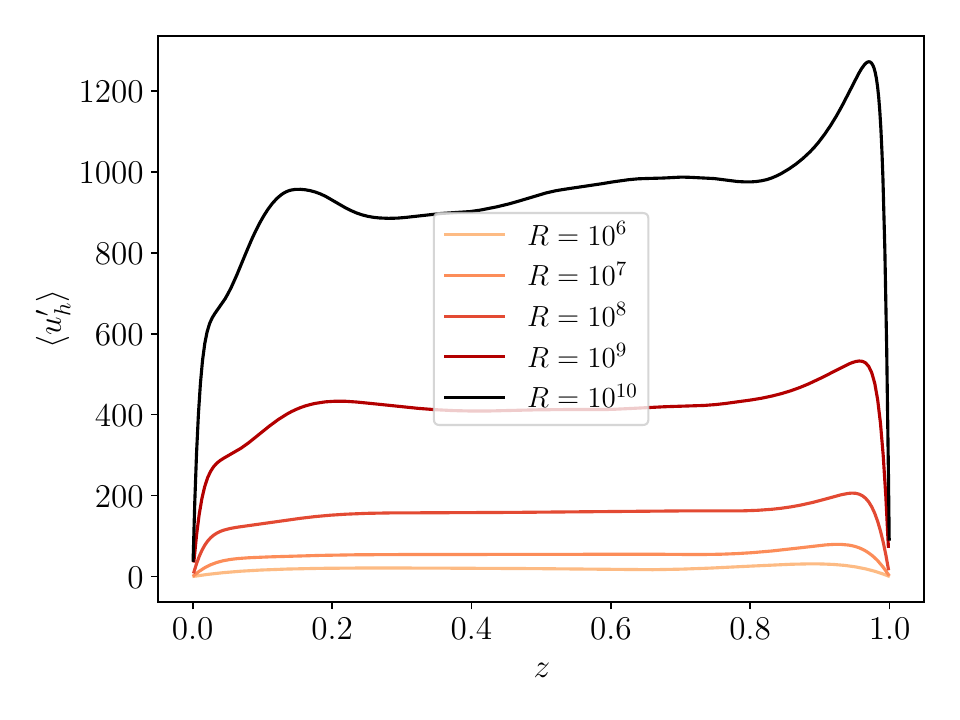}
\includegraphics[width=.45\linewidth]{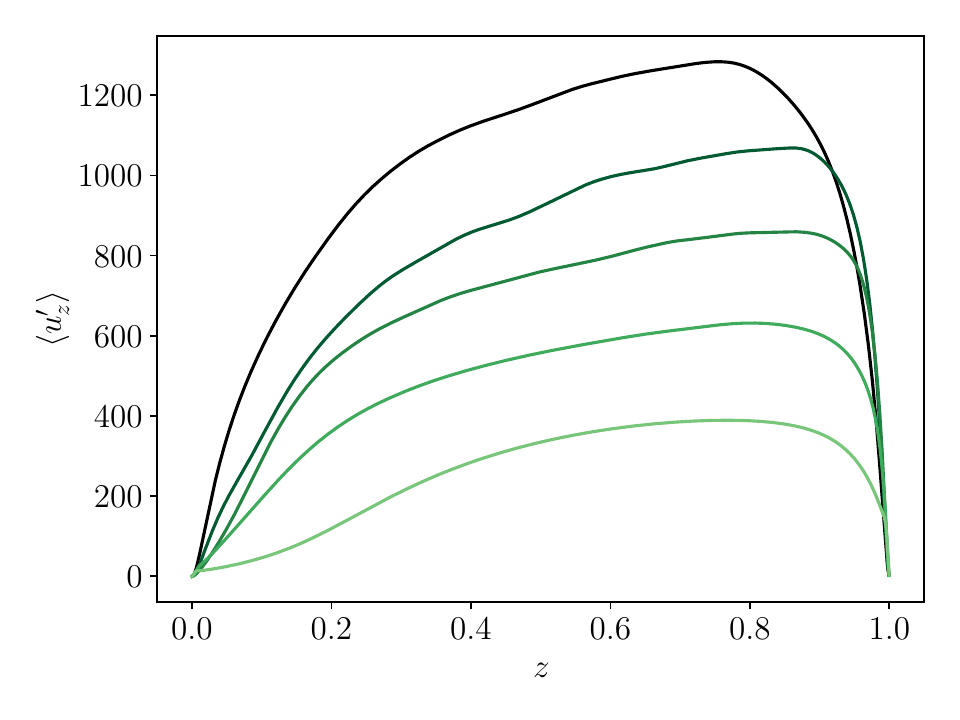}
\includegraphics[width=.45\linewidth]{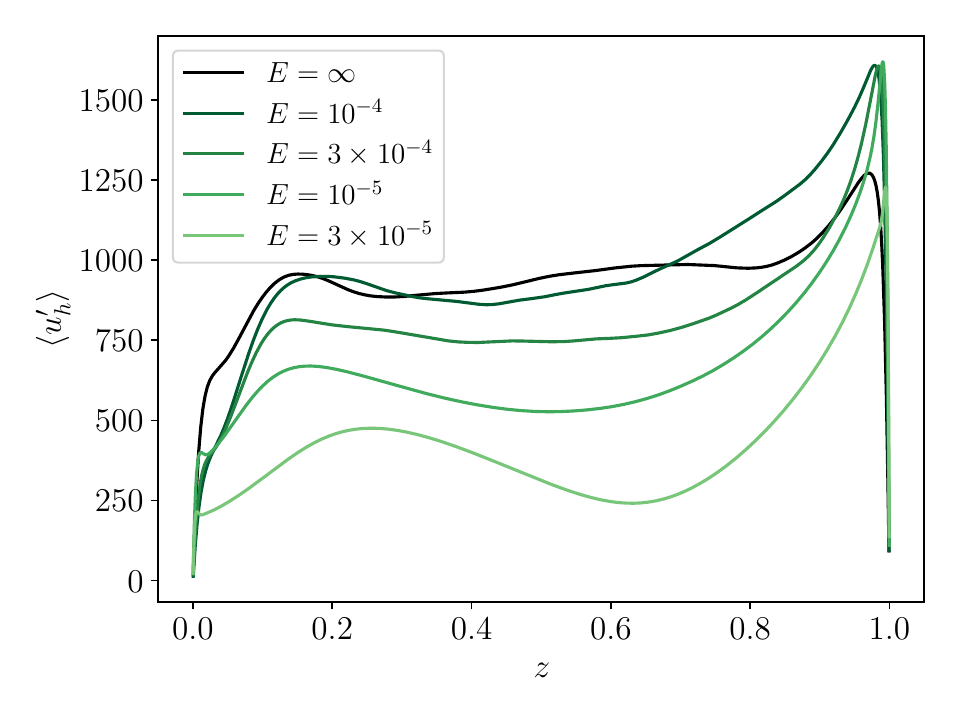}
\caption{Velocity fluctuations in the vertical (left column) and horizontal (right column) directions for several values of $R$ in the non-rotating case (top row) and for several values of $E$ at $R=10^{10}$ (bottom row).  }
\label{fig:ustats}
\end{figure}

We now examine the velocity statistics. To quantify the fluctuations in the horizontal direction we define the horizontal velocity $u_h=\sqrt{u_x^2+u_y^2}$. Figure \ref{fig:ustats} presents both the horizontal $\langle u_h^\prime \rangle$ and vertical $\langle u_z^\prime\rangle$ velocity fluctuations. The vertical velocity follows a straightforward trend aligning with expectations: fluctuations (and transport) increase with $z$ before decreasing near the velocity boundary layer at the top plate. When rotation is introduced, vertical velocity is increasingly suppressed as $E$ decreases. However, the overall shape of the profile remains relatively unchanged, aside from the fact that at the smallest $E$, the stably stratified layer becomes increasingly quiet.

In contrast, horizontal motions exhibit more complex behaviour. In the absence of rotation, the upper half of the system behaves similarly to RBC: fluctuations peak near the wall, marking the edge of the boundary layer, while the bulk region maintains an almost uniform fluctuation level. Introducing rotation preserves this overall structure but thins the boundary layer and enhances fluctuations as it transitions into an Ekman layer.

The lower half of the domain features two distinct local maxima in fluctuations. The peak closest to the wall corresponds to the velocity boundary layer, while the second peak is caused by the interface between stable and unstably stratified regions. This peak roughly aligns with the position of the minima in $\langle T^\prime\rangle$ discussed earlier. As rotation increases, these effects become more pronounced: the first peak shifts closer to the wall as the velocity boundary layer becomes a thin Ekman layer, while the second peak moves outwards as reduced mixing allows the stably stratified layer to extend further into the flow. This showcases the presence of two distinct regions near the bottom plate- the velocity boundary layer and the stably stratified region, and that their interaction is intricate.

\subsection{Boundary layer behaviour}
\label{sec:bls}

\begin{figure}
\centering
\includegraphics[width=.45\linewidth]{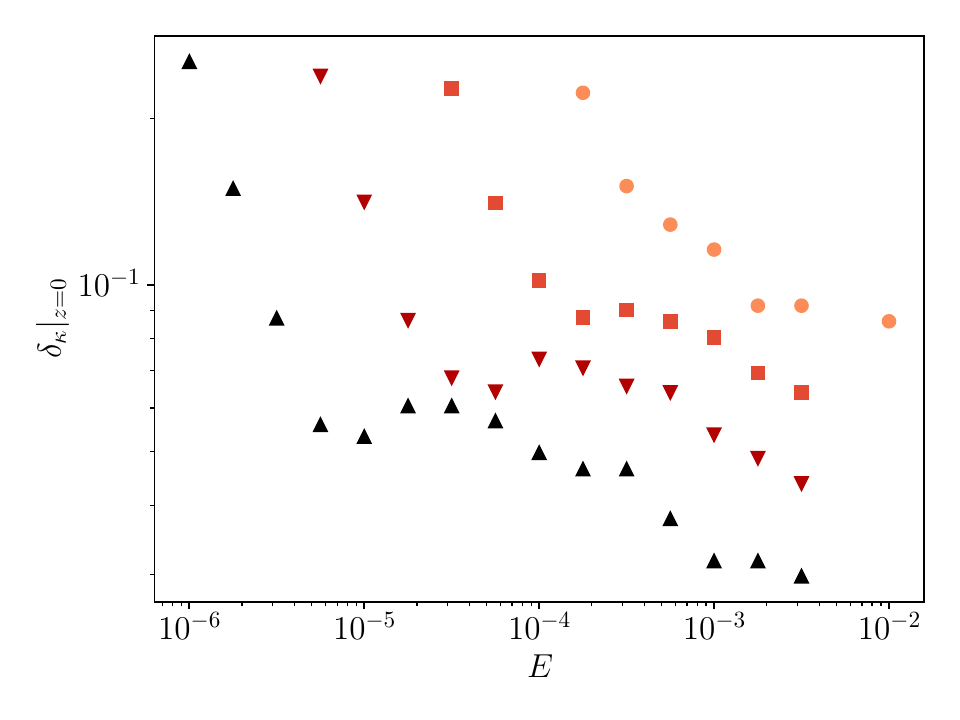}
\includegraphics[width=.45\linewidth]{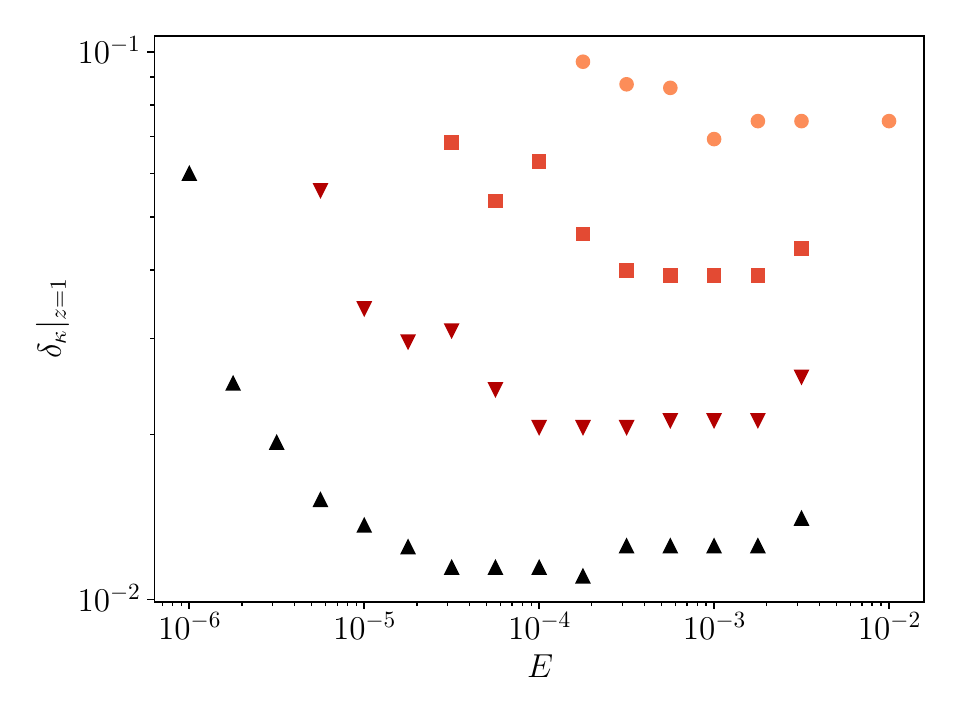}
\caption{Extent of the thermal boundary layers $\delta_\kappa$ at the bottom (left) and top (right) plates. }
\label{fig:dTsize1}
\end{figure}

Before examining the exact relationships for the dissipation, the extent and type of the boundary layers present in the system must be analysed. We begin with the thermal boundary layers whose thickness $\delta_\kappa$ is determined using the peak in the $\langle T^\prime \rangle$ profiles. Figure \ref{fig:dTsize1} illustrates the dependence of $\delta_\kappa$ on $R$ and $E$. In general, increasing $R$ leads to thinner thermal boundary layers due to enhanced convection. The bottom boundary layer extends further into the flow compared to the top one, as stable stratification inhibits heat transport. While rotation generally increases the boundary layer thickness, some non-monotonic behaviour is observed, likely resulting from the competing effects of rotational suppression of turbulence and improved flow organisation, which enhances heat transport.

\begin{figure}
\centering
\includegraphics[width=.45\linewidth]{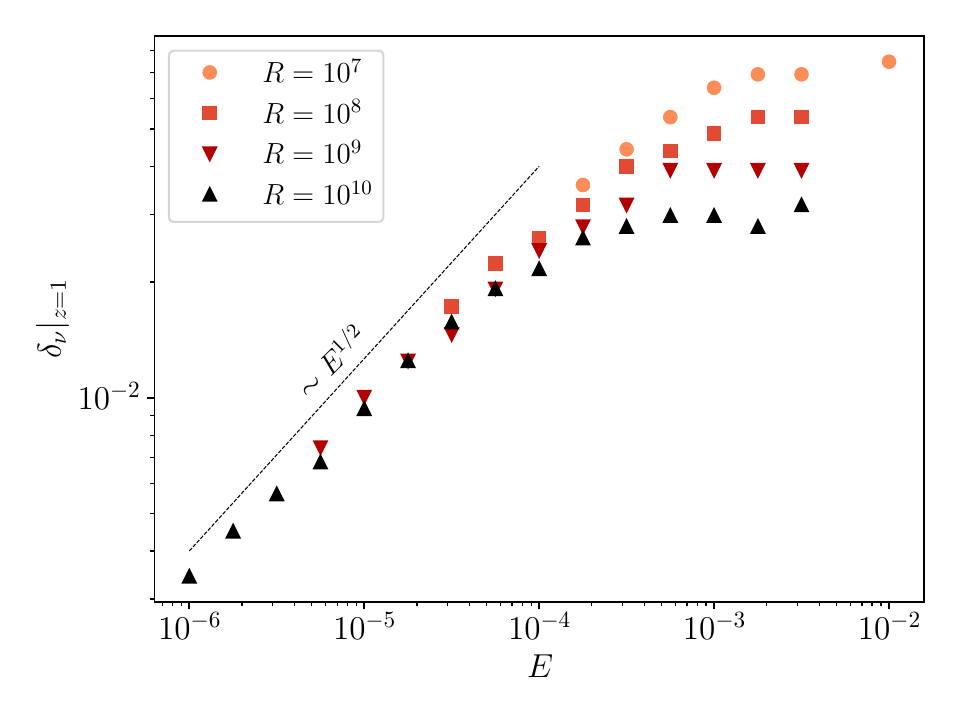}
\includegraphics[width=.45\linewidth]{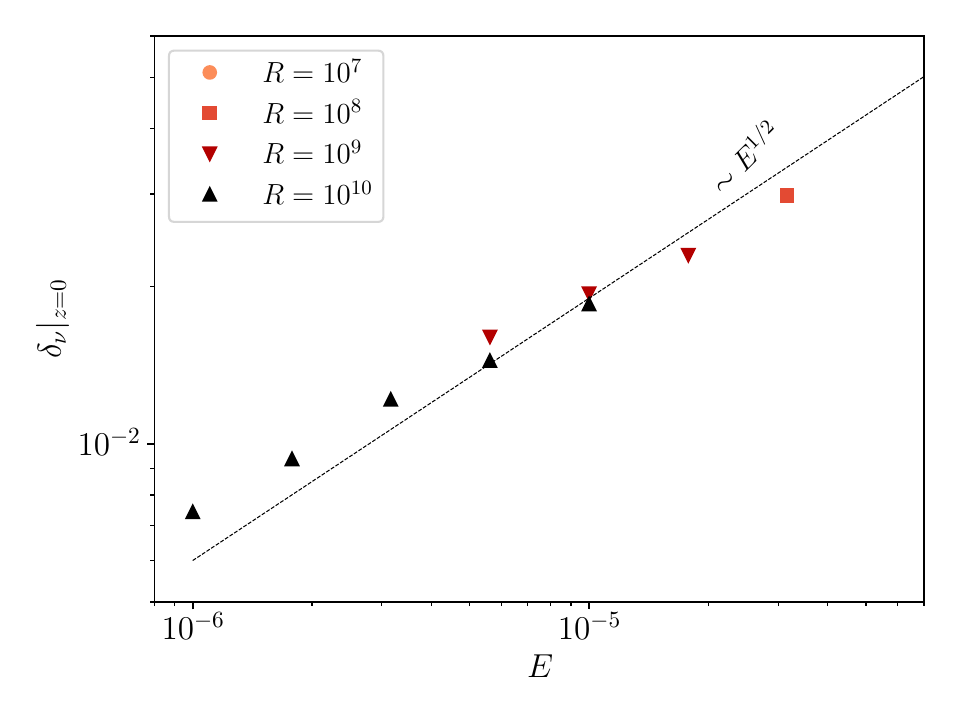}
\caption{Extent of the top (left) and bottom (right) velocity boundary layer $\delta_\nu|$ as a function of $E$. Symbols as in Figure \ref{fig:tgrad}. }
\label{fig:ekbls}
\end{figure}

Turning to the velocity boundary layers, the top boundary layer behaves in a straightforward manner. Its thickness can be defined from the peak in $\langle u_h^\prime \rangle$, with its dependence on $E$ and $R$ shown in the left panel of Figure \ref{fig:ekbls}. At high rotation rates, the data show reasonable collapse when plotted as a function of $E$, following the expected scaling $\delta_\nu|_{z=1}\sim\mathcal{O}(E^{1/2})$. For sufficiently small $E$, the dependence on $R$ drops out as expected

The bottom boundary layer, however, is more difficult to quantify, as the fluctuation peak only becomes distinct at sufficiently small $E$. As a result, its thickness can only be precisely determined once the boundary layer has transitioned into an Ekman layer. In the right panel of Figure \ref{fig:ekbls}, we present measurements of the bottom boundary layer extent for cases where the peak is clearly identifiable. The available data points are limited in quantity, but we can again observe a scaling similar to $\delta_\nu|_{z=0}\sim\mathcal{O}(E^{1/2})$, with minimal $R$ dependence at small $E$. 

\begin{figure}
\centering
\includegraphics[width=.45\linewidth]{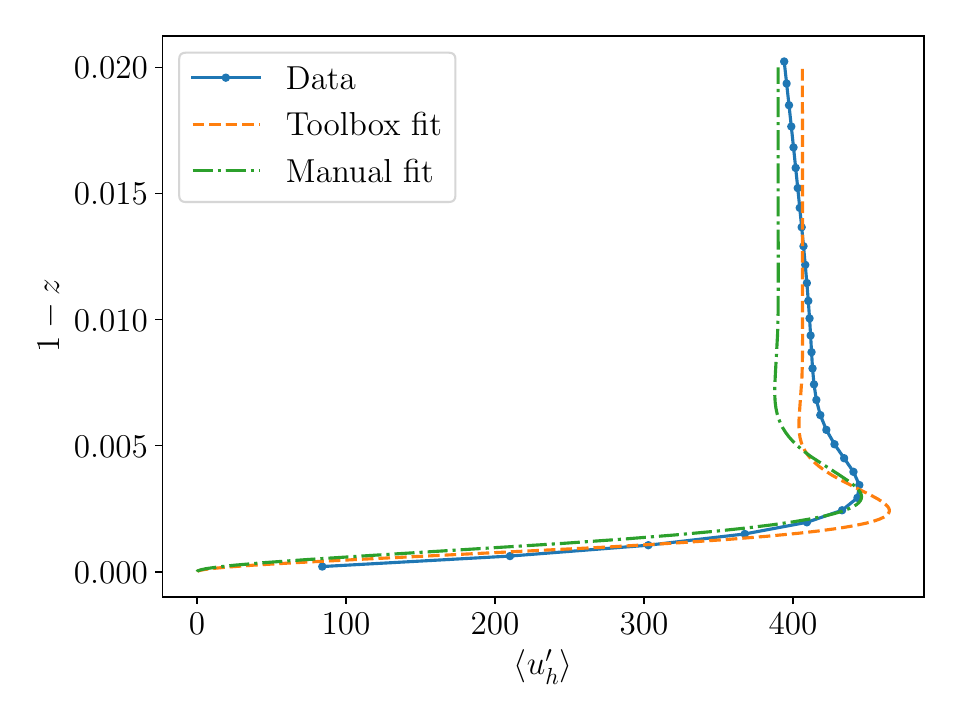}
\includegraphics[width=.45\linewidth]{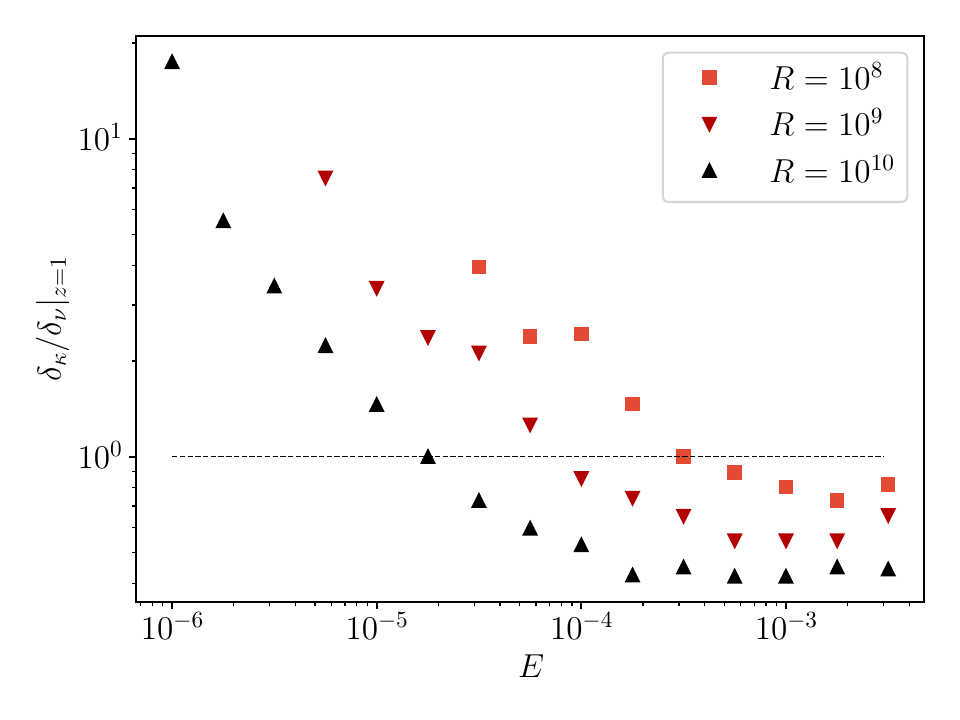}
\caption{Left: Numerically obtained velocity profile for $R=10^{10}$ and $E=10^{-6}$ compared to the theoretical Ekman solution fitted manually and using a least-squares fit. Right: Size ratio between thermal and velocity boundary layers at the top plate. }
\label{fig:dTsize}
\end{figure}

It is important to note that these boundary layers are not sufficiently thin to exhibit ``pure'' Ekman dynamics, as seen in rotating RBC at small $E$. In the left panel of Figure \ref{fig:dTsize}, we follow \cite{aguirre2022flow} and attempt to fit the theoretical solution for the Ekman boundary layer to the profiles obtained. In general, we cannot reproduce the peak, using both a least-squares fit or a manual fit which captures the size of the peak but fails to capture the rest of the behaviour. 

Finally, the right panel of Figure \ref{fig:dTsize} presents the ratio of thermal to velocity boundary layer thicknesses at the top plate. This ratio has been proposed as a diagnostic for the transition to rotation-dominated convection or even the onset of geostrophic turbulence \citep{king2009boundary,king2012thermal}. However, in our simulations, this crossover occurs at relatively large values of $E$ (i.e., weak rotation) compared to RBC. This supports the argument by \cite{kunnen2016transition} that this ratio alone does not govern the transition to rotation-dominated convection.

\subsection{Dissipation rates}
\label{sec:diss}

Having described the phenomenology of the flow, we conclude our analysis by examining the dissipation rates $\epsilon_\nu$ and $\epsilon_\theta$. These quantities are related to the global quantities $\wT$ and $\T$ via the equations \ref{eq:exrel1}-\ref{eq:exrel2}. Investigating their local behaviour and the relative  contributions from the bulk and boundary layers has been a key approach in RBC to estimate the global response of the system and identify the factors limiting heat transport.

\begin{figure}
\centering
\includegraphics[width=.45\linewidth]{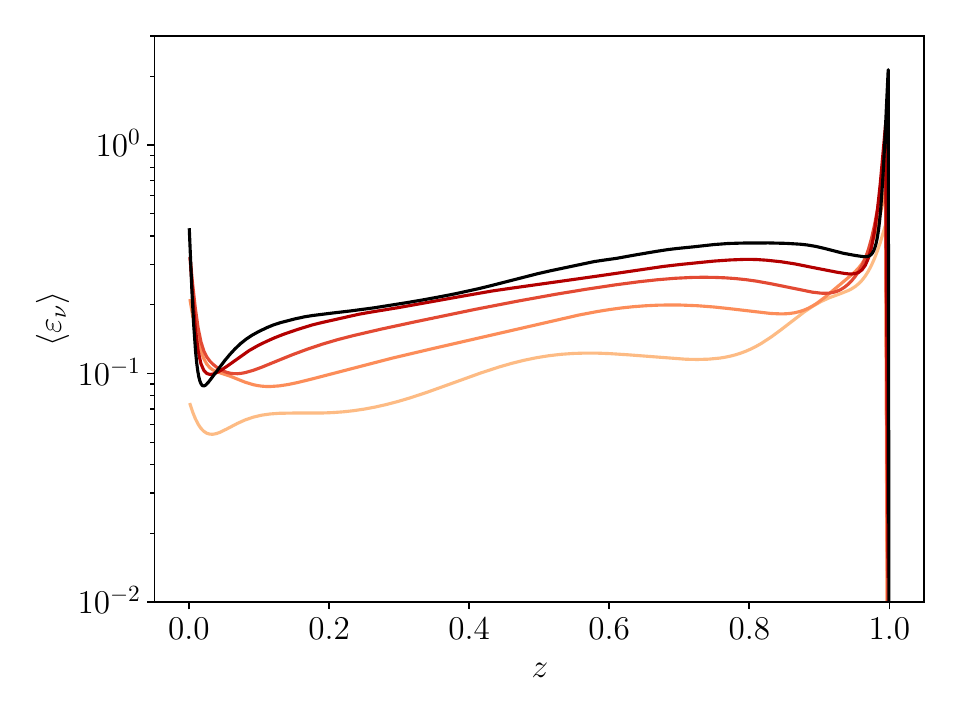}
\includegraphics[width=.45\linewidth]{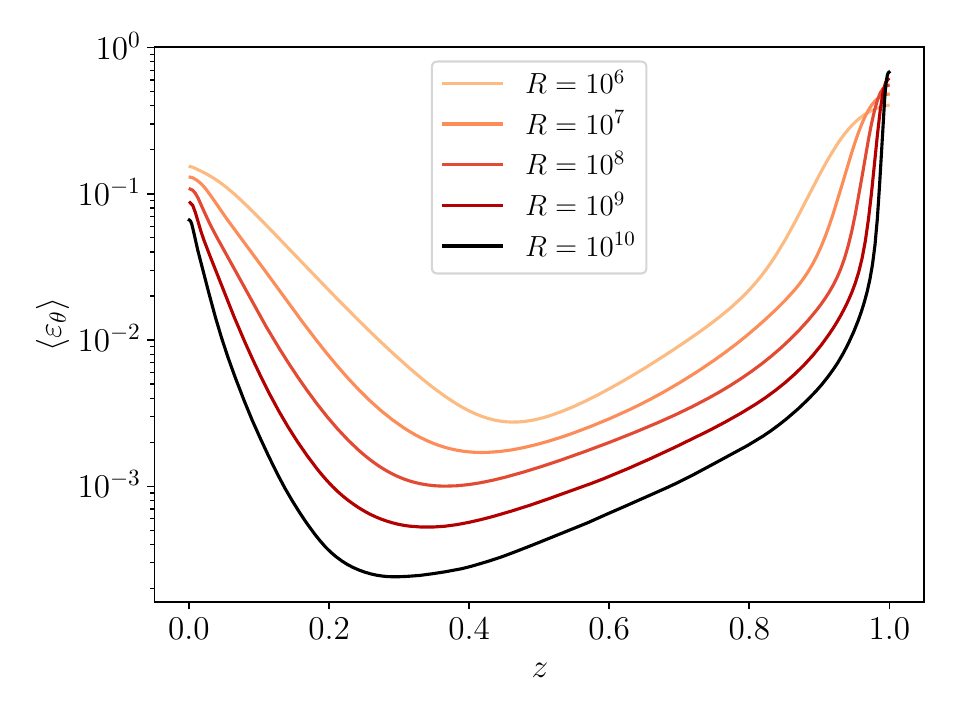}
\includegraphics[width=.45\linewidth]{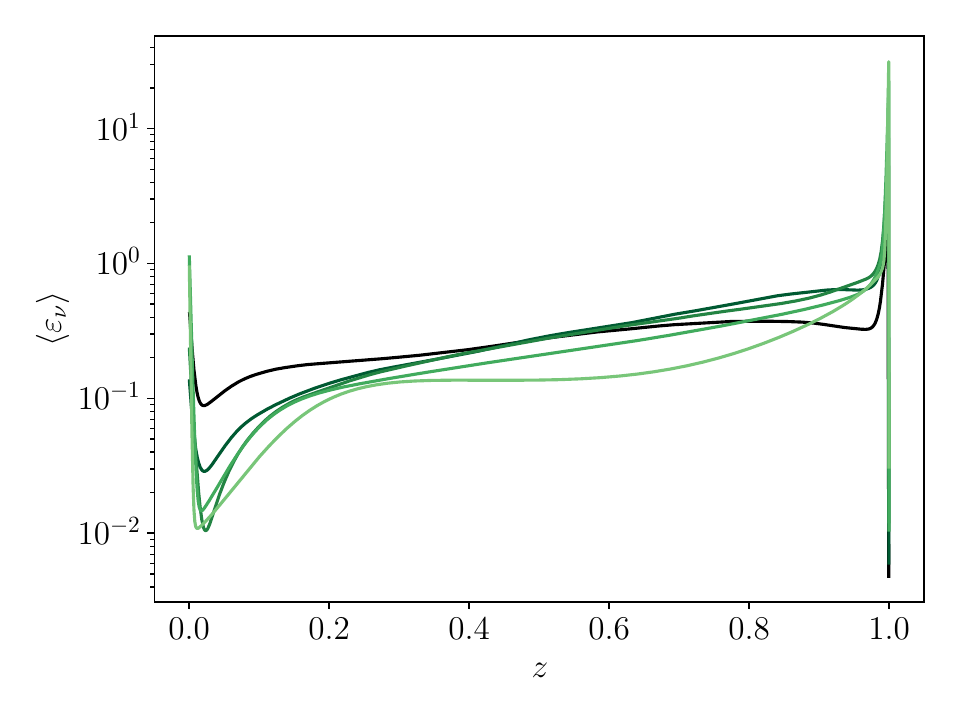}
\includegraphics[width=.45\linewidth]{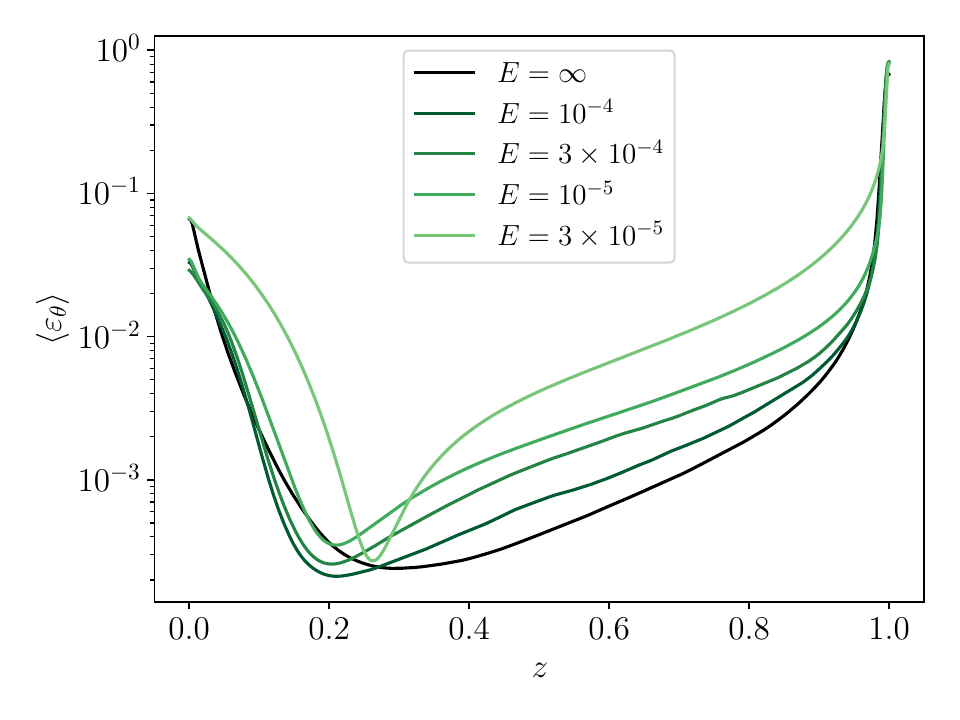}
\caption{Profiles of $\langle \epsilon_\nu \rangle$ (left column) and $\langle \epsilon_\theta \rangle$ (right column) dissipation rates for several values of $R$ for the non-rotating case (top row) and several values of $E$ for $R=10^{10}$ (bottom row). }
\label{fig:epsprofiles}
\end{figure}

The top row of Figure \ref{fig:epsprofiles} shows the profiles of $\langle\epsilon_\nu\rangle$ for various values of $R$ and $E$. Generally, these profiles feature a bulk region with nearly constant values of $\epsilon_\nu$ and near-wall regions with local maxima and minima. At the top wall, kinetic energy dissipation shows a peak, consistent with expectations. However, the bottom wall displays a more complex structure: a peak at the wall, and a local minima just beyond the velocity boundary layer. As rotation increases, this minimum becomes more pronounced, while the dissipation peak at the top boundary layer strengthens due to the intensified kinetic energy dissipation within the Ekman boundary layer.

The bottom row of Figure \ref{fig:epsprofiles} shows the profiles of $\langle\epsilon_\theta\rangle$ for the same values of $R$ and $E$. Unlike $\epsilon_\nu$, $\epsilon_\theta$ decreases with increasing $R$, reflecting the overall reduction in temperature fluctuations. Local maxima appear across all cases at the plates, and a distinct minimum can be observed farther the bottom wall. The location of this minimum roughly corresponds to that of $\langle T^\prime\rangle$ seen in Figure \ref{fig:tprofiles}, which we previously associated with the edge of the stably stratified layer. Increasing rotation does not significantly alter the shape of the $\epsilon_\theta$ profiles or the magnitude of the peak at the top plate, though the position of the local minimum shifts slightly.

Following \cite{grossmann2000scaling, Wang2020}, we decompose the dissipation rates $\epsth$ into contributions from the bulk and boundary layers to better understand the scaling laws governing $\T$ and $\wT$. Defining the bottom velocity boundary layer extent is challenging, preventing us from isolating its contribution to $\epsth$. However, inspection of the $\langle\epsilon_\nu\rangle$ profiles in Figure \ref{fig:epsprofiles} suggests that dissipation in the bottom boundary layer is negligible compared to the bulk contribution, justifying the assumption that a combined bulk - bottom BL contributions would primarily originate from the bulk region.

\begin{figure}
\centering
\includegraphics[width=.45\linewidth]{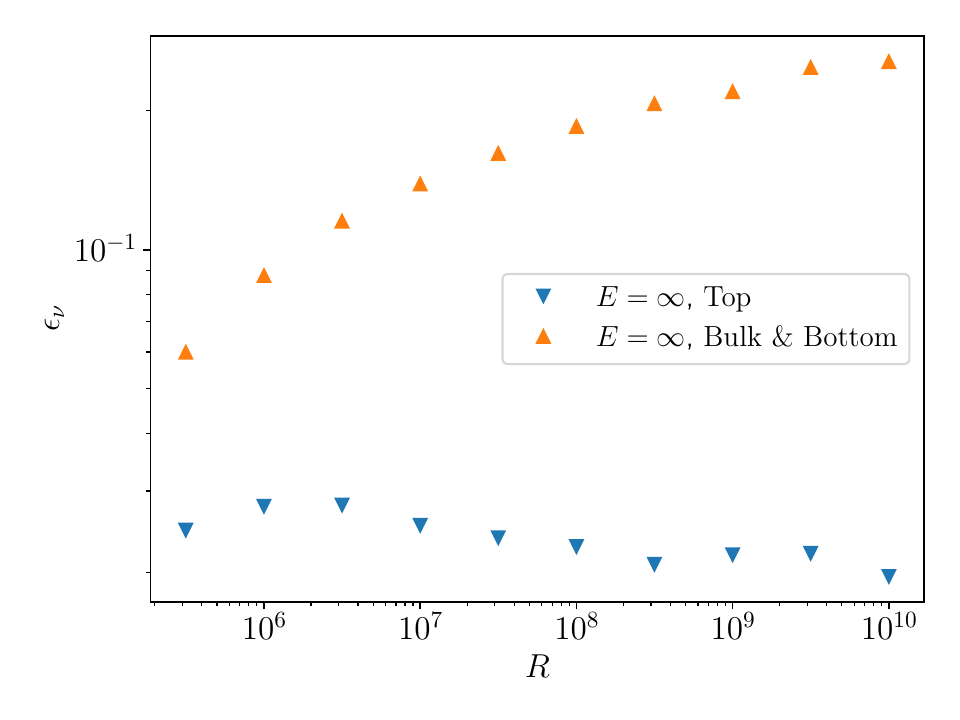}
\includegraphics[width=.45\linewidth]{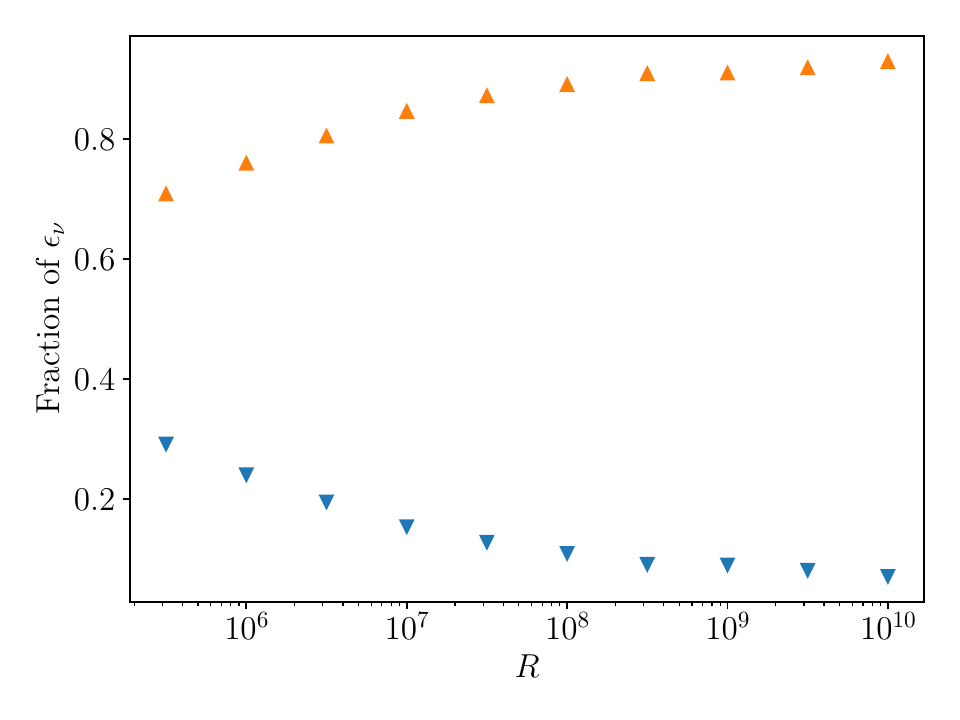}
\includegraphics[width=.45\linewidth]{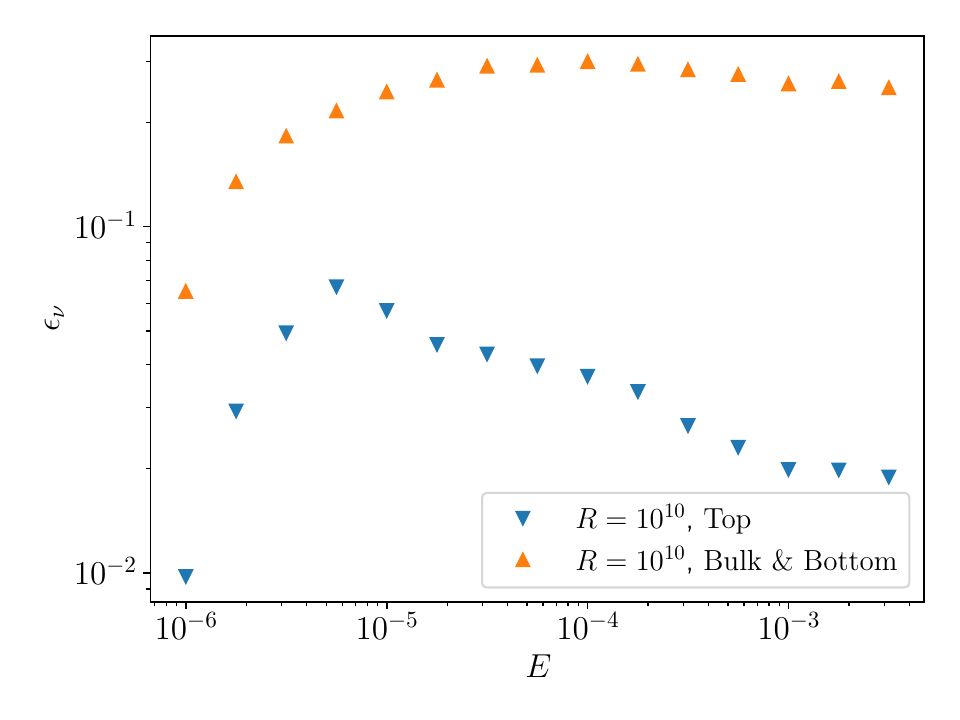}
\includegraphics[width=.45\linewidth]{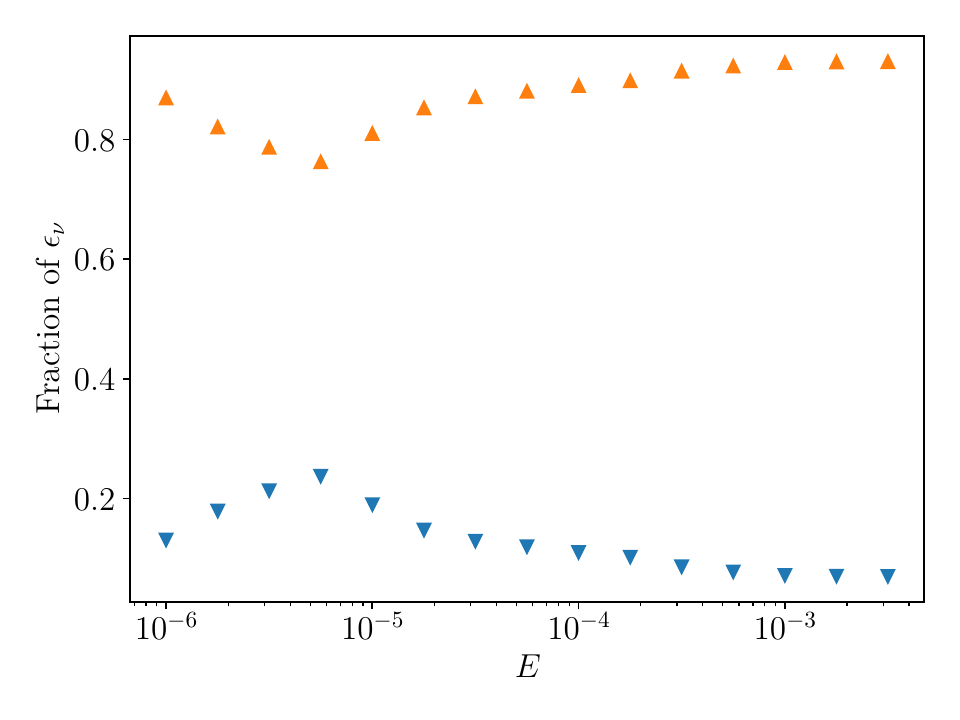}
\caption{Absolute (left) and relative (right) contributions to the viscous dissipation rate $\epsnu$ for non-rotating IHC (top row) and rotating IHC at $R=10^{10}$ (bottom row). }
\label{fig:epsnucont}
\end{figure}

Figure \ref{fig:epsnucont} shows the absolute and relative contributions of $\epsnu$ from the top boundary and from the bulk (including the bottom boundary layer). In non-rotating cases, $\epsnu$ increases in the bulk with increasing $R$, but decreases at the top BL. The relative contributions indicate that over $80\%$ of $\epsnu$ comes from the bulk, suggesting that the dependence of $\wT$ on $R$ is primarily governed by bulk viscous dissipation.

With increasing rotation, the behaviour becomes more complex. While $\epsnu$ in the bulk remains nearly constant before decreasingly rapidly at small values of $E$, dissipation in the top boundary layer increases significantly, reaching relative contributions of over $20\%$. The maxima in $\wT(E)$ can be directly linked to increases in $\epsnu$ within the top boundary layer, preceding the decline of $\epsnu$ in the bulk. 

\begin{figure}
\centering
\includegraphics[width=.45\linewidth]{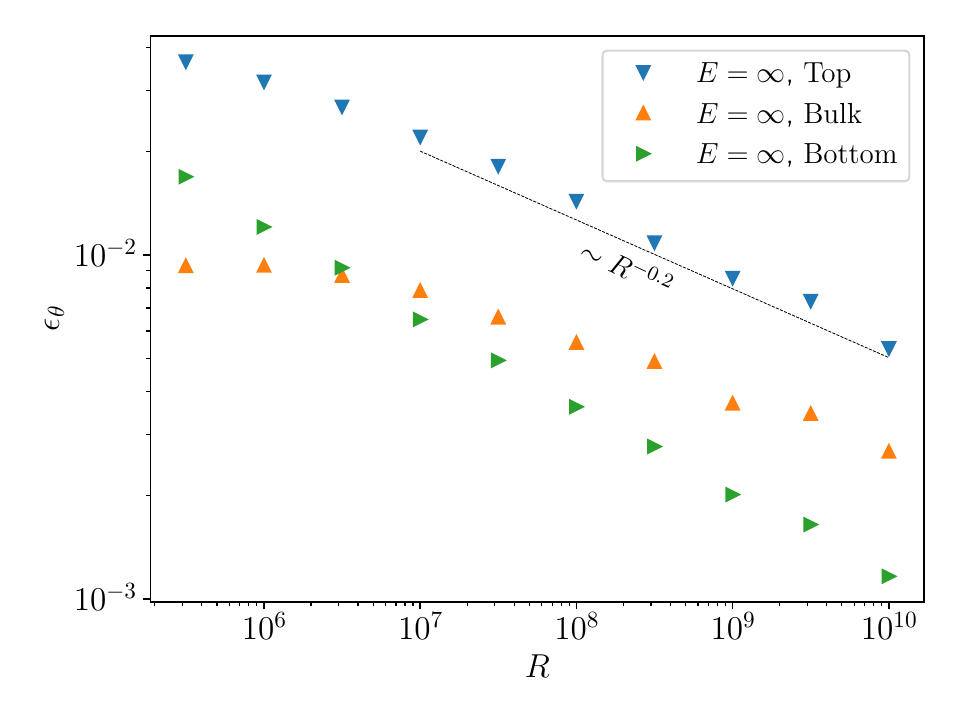}
\includegraphics[width=.45\linewidth]{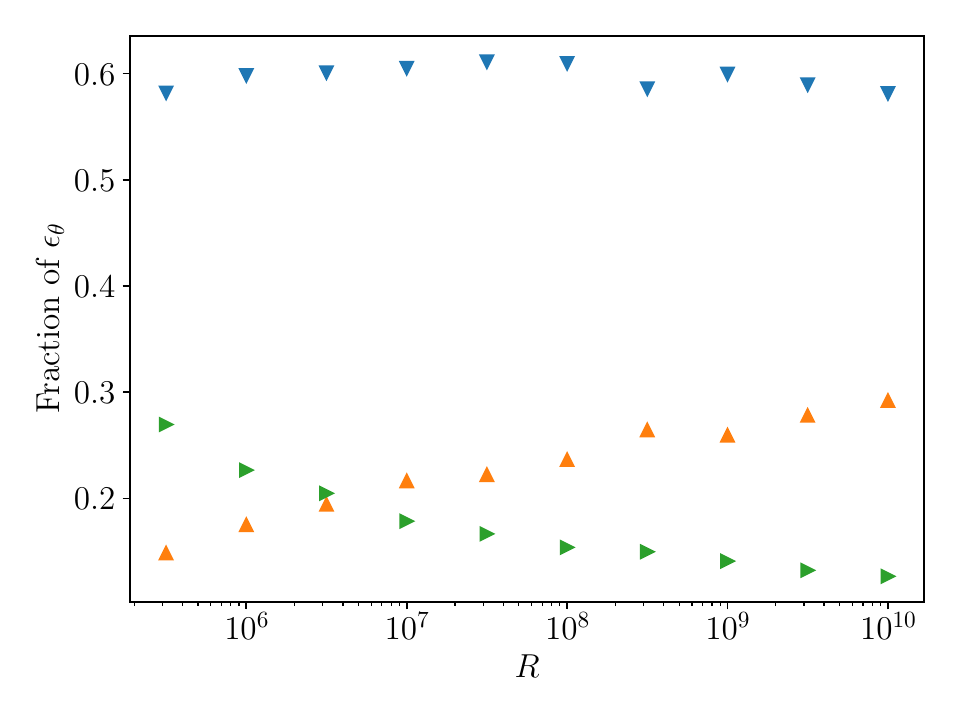}
\includegraphics[width=.45\linewidth]{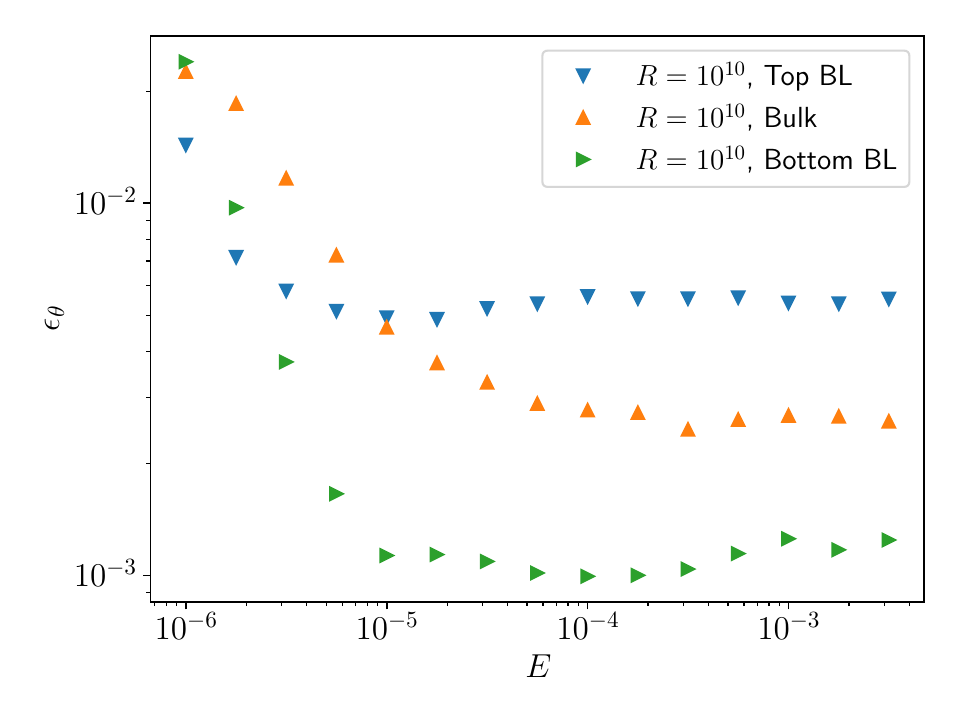}
\includegraphics[width=.45\linewidth]{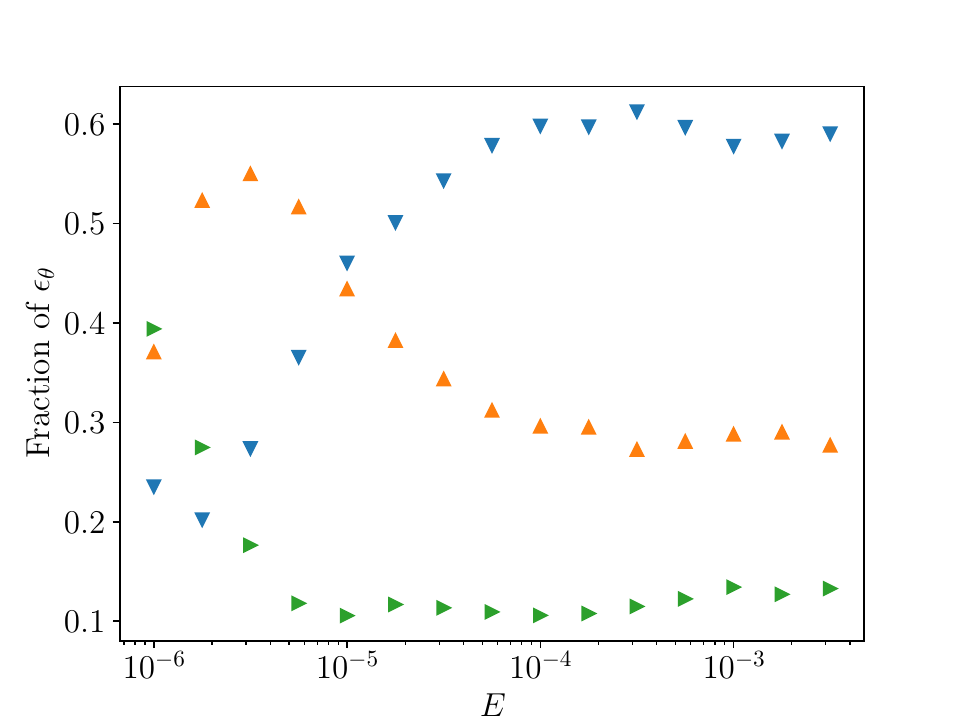}
\caption{Absolute (left) and relative (right) contributions to the thermal dissipation rate $\epsth$ for non-rotating IHC (top row) and rotating IHC at $R=10^{10}$ (bottom row). }
\label{fig:epsthcont}
\end{figure}

A similar analysis for $\epsth$ is shown in Figure \ref{fig:epsthcont}. In the absence of rotation, the dominant contribution comes from the top boundary layer. A rough scaling of $R^{-0.2}$ is observed for thermal dissipation in this region, indicating that the scaling laws governing $\T(R)$ are constrained by flow dynamics in this region. As $R$ increases, bulk contributions become increasingly significant, suggesting that for much larger values of $R$, the bulk may eventually dominate the behaviour of $\T$.

The behaviour of $\epsth$ under rotation differs significantly from that of $\epsnu$. While dissipation in the top thermal boundary layers decreases slightly at high rotation rates, this effect is quickly overshadowed by a surge in bulk dissipation, which surpasses boundary layer contributions and limits the decrease of $\T$. This is another explanation on why $\T$ and $\wT$ exhibit different trends around ``optimal transport'' from a dissipation perspective, and why $\T$ is more strongly dependent on $R$, as seen from Figure \ref{fig:opttrans}. 

Our findings indicate that for $Pr=1$, $\epsnu$ is primarily controlled by bulk dissipation, wheareas $\epsth$ is dominated by dissipation in the top boundary layer. This contrasts with the 2D simulations of \cite{Wang2020}, which suggest that $\epsnu$ is primarily governed by boundary layer contributions. The discrepancy likely arises from the behavior of internally heated convection (IHC) in two dimensions, where the presence of a large-scale circulation significantly biases velocity statistics, causing the velocities to be higher and hence resulting in much larger dissipations \citep{goluskin2016penetrative}.

\section{Conclusion and Outlook}
\label{sec:conc}

In this study, we have conducted the first three dimensional simulations of rotating internally heated convection with isothermal boundaries, where the fluid is cooled from both plates. We demonstrated that the asymptotic stability results derived in \cite{arslan2024internally} accurately describe the behaviour observed in our simulations at $Pr=1$. We then explored the effects of rotation on IHC over a parameter range of $R \in [3.16\times10^5,10^{10}]$ and $E \in [10^{-6},\infty)$. Weak rotation organises the plumes originating from the top boundary layer, enhancing heat transport even as the velocity fluctuations are reduced. This effect is reflected on both global quantities $\T$ and $\wT$, with a more pronounced impact on the latter. We also analysed local flow behaviour, revealing that an Ekman boundary layer forms near the top plate, while near the bottom plate, a complex interaction occurs between the boundary layers and the stably stratified region. This effect is unique to IHC in the IH1 configuration, and is affected by rotation. Finally, we examined the dissipation rates, concluding that the system was dominated by boundary layer dissipation for $\epsth$ and bulk dissipation for $\epsnu$, a significant difference with the two-dimensional simulations.

This study is has two main limitations. First, we have fixed $Pr=1$, and further exploration of different Prandtl numbers would provide deeper insight into the system’s complex dynamics, especially into the interplay between boundary layer(s) and stably stratified regions. System asymmetry has been shown to substantially depend on $Pr$ \citep{goluskin2016penetrative,Wang2020}. Second, our thermal driving is not strong enough to access the fully rotation-dominated regime, where $E$ can be varied without significantly suppressing convection. Future simulations extending into this regime will be essential for a more comprehensive understanding of rotating IHC. In an extended parameter range the sharpness of rigorous bounds can be further tested and possibly new flow states can be probed for rotating IHC.   
\\

\noindent {\it Acknowledgments:} ROM acknowledges support from the Emergia Program of the Junta de Andalucía (Spain).
AA acknowledges funding from the ERC (agreement no. 833848-UEMHP) under the Horizon 2020 program and the SNSF (grant number 219247) under the MINT 2023 call. The authors thank Rudie Kunnen for helpful comments. We also thank the Systems Unit of the Information Systems Area of the University of Cadiz for computer resources and technical support.

\noindent {\it Declaration of Interests:} The authors report no conflict of interest.

\appendix

\section{Summary of results}
\label{sec:results}

Table \ref{tbl:summary} presents a summary of all numerical results.

\afterpage{\clearpage}
\begin{landscape}
\begin{longtable}[c]{|c|c|c|c|c|c|c|c|c|c|}
\caption{Summary of results}
\label{tbl:summary}\\

\hline
$R$ & $Ek$ & $\tilde{R}$ & $Ro$ & $\Gamma$ & $N_{x,y}\times N_z$ & $\langle T\rangle$ & $\langle wT \rangle$ & $Re_w$ \\
\hline
\endhead

$3.16\times 10^5$ & $\infty$ & - & - & $3.14$ & $288^2\times144$ & 
$6.43\times10^{-2}$ & $8.68\times 10^{-2}$ & $1.90\times 10^{1}$ \\
$3.16\times 10^5$ & $1.00\times10^{-2}$ & $6.81\times10^2$ & $5.62\times10^0$ & $3.14$ & $288^2\times144$  & 
$6.46\times10^{-2}$ & $8.74\times 10^{-2}$ & $1.78\times 10^{1}$ \\
$3.16\times 10^5$ & $3.16\times10^{-3}$ & $1.47\times10^2$ & $1.78\times10^0$ & $3.14$ & $288^2\times144$  & 
$7.09\times10^{-2}$ & $5.46\times 10^{-2}$ & $1.15\times 10^{1}$ \\
$3.16\times 10^5$ & $1.78\times10^{-3}$ & $6.83\times10^1$ & $1.00\times10^0$ & $3.14$ & $288^2\times144$  & 
\multicolumn{3}{c|}{Fully conductive} \\
\hline
$1.00\times 10^6$ & $\infty$ & - & - & $3.14$ & $288^2\times144$ & 
$5.48\times10^{-2}$ & $1.22\times 10^{-1}$ & $3.42\times 10^{1}$ \\
$1.00\times 10^6$ & $1.00\times10^{-2}$ & $2.15\times10^3$ & $1.00\times10^1$ & $3.14$ & $288^2\times144$  & 
$5.46\times10^{-2}$ & $1.24\times 10^{-1}$ & $3.39\times 10^{1}$ \\
$1.00\times 10^6$ & $3.16\times10^{-3}$ & $4.65\times10^2$ & $3.16\times10^0$ & $3.14$ & $288^2\times144$  & 
$5.53\times10^{-2}$ & $1.25\times 10^{-1}$ & $2.90\times 10^{1}$ \\
$1.00\times 10^6$ & $1.78\times10^{-3}$ & $2.16\times10^2$ & $1.78\times10^0$ & $3.14$ & $288^2\times144$  & 
$5.90\times10^{-2}$ & $1.02\times 10^{-1}$ & $2.51\times 10^{1}$ \\
$1.00\times 10^6$ & $1.00\times10^{-3}$ & $1.00\times10^2$ & $1.00\times10^0$ & $3.14$ & $288^2\times144$  & 
$7.03\times10^{-2}$ & $5.16\times 10^{-2}$ & $1.42\times 10^{1}$ \\
$1.00\times 10^6$ & $5.00\times10^{-4}$ & $3.97\times10^1$ & $5.00\times10^{-1}$ & $3.14$ & $288^2\times144$  & 
\multicolumn{3}{c|}{Fully conductive} \\
\hline 
$3.16\times 10^6$ & $\infty$ & - & - & $3.14$ & $288^2\times144$ & 
$4.60\times10^{-2}$ & $1.47\times 10^{-1}$ & $5.80\times 10^{1}$ \\
$3.16\times 10^6$ & $1.00\times10^{-2}$ & $6.81\times10^3$ & $1.78\times10^2$ & $3.14$ & $288^2\times144$  & 
$4.61\times10^{-2}$ & $1.47\times 10^{-1}$ & $5.73\times 10^{1}$ \\
$3.16\times 10^6$ & $3.16\times10^{-3}$ & $1.47\times10^3$ & $5.63\times10^0$ & $3.14$ & $288^2\times144$  & 
$4.51\times10^{-2}$ & $1.67\times 10^{-1}$ & $5.56\times 10^{1}$ \\
$3.16\times 10^6$ & $1.00\times10^{-3}$ & $3.16\times10^2$ & $1.78\times10^0$ & $3.14$ & $288^2\times144$  & 
$4.84\times10^{-2}$ & $1.49\times 10^{-1}$ & $5.80\times 10^{1}$ \\
$3.16\times 10^6$ & $5.62\times10^{-4}$ & $4.21\times10^2$ & $1.00\times10^0$ & $3.14$ & $288^2\times144$  & 
$5.68\times10^{-2}$ & $9.97\times 10^{-2}$ & $2.98\times 10^{1}$ \\
$3.16\times 10^6$ & $3.16\times10^{-4}$ & $6.81\times10^1$ & $5.63\times10^{-1}$ & $3.14$ & $288^2\times144$  & 
$7.74\times10^{-2}$ & $2.05\times 10^{-2}$ & $1.12\times 10^{1}$ \\
$3.16\times 10^6$ & $1.78\times10^{-4}$ & $3.17\times10^1$ & $3.17\times10^{-1}$ & $3.14$ & $288^2\times144$  & 
\multicolumn{3}{c|}{Fully conductive} \\
\hline 
$1.00\times 10^7$ & $\infty$ & - & - & $3.14$ & $288^2\times144$ & 
$3.82\times10^{-2}$ & $1.67\times 10^{-1}$ & $9.41\times 10^{1}$ \\
$1.00\times 10^7$ & $1.00\times10^{-2}$ & $2.15\times10^4$ & $3.61\times10^1$ & $3.14$ & $288^2\times144$  & 
$3.83\times10^{-2}$ & $1.70\times 10^{-1}$ & $9.46\times 10^{1}$ \\
$1.00\times 10^7$ & $3.16\times10^{-3}$ & $4.65\times10^3$ & $1.00\times10^1$ & $3.14$ & $288^2\times144$  & 
$3.81\times10^{-2}$ & $1.87\times 10^{-1}$ & $9.52\times 10^{1}$ \\
$1.00\times 10^7$ & $1.78\times10^{-3}$ & $2.16\times10^3$ & $5.63\times10^0$ & $3.14$ & $288^2\times144$  & 
$3.77\times10^{-2}$ & $1.97\times 10^{-1}$ & $9.21\times 10^{1}$ \\
$1.00\times 10^7$ & $1.00\times10^{-3}$ & $1.00\times10^3$ & $3.16\times10^0$ & $3.14$ & $288^2\times144$  & 
$3.69\times10^{-2}$ & $2.03\times 10^{-1}$ & $8.26\times 10^{1}$ \\
$1.00\times 10^7$ & $5.62\times10^{-4}$ & $4.64\times10^2$ & $1.78\times10^0$ & $3.14$ & $288^2\times144$  & 
$3.88\times10^{-2}$ & $1.92\times 10^{-1}$ & $7.15\times 10^{1}$ \\
$1.00\times 10^7$ & $3.16\times10^{-4}$ & $2.16\times10^2$ & $1.00\times10^0$ & $3.14$ & $288^2\times288$  & 
$4.56\times10^{-2}$ & $1.49\times 10^{-1}$ & $5.48\times 10^{1}$ \\
$1.00\times 10^7$ & $1.78\times10^{-4}$ & $1.00\times10^2$ & $5.64\times10^{-1}$ & $3.14$ & $288^2\times288$  & 
$6.20\times10^{-2}$ & $7.32\times 10^{-2}$ & $3.13\times 10^{1}$ \\
$1.00\times 10^7$ & $1.00\times10^{-4}$ & $4.68\times10^1$ & $3.16\times10^{-1}$ & $3.14$ & $288^2\times288$  & 
\multicolumn{3}{c|}{Fully conductive} \\
\hline
$3.16\times 10^7$ & $\infty$ & - & - & $3.14$ & $288^2\times144$ & 
$3.15\times10^{-2}$ & $1.88\times 10^{-1}$ & $1.54\times 10^{2}$ \\
$3.16\times 10^7$ & $3.16\times10^{-3}$ & $1.47\times10^4$ & $1.78\times10^1$ & $3.14$ & $288^2\times144$  & 
$3.16\times10^{-2}$ & $2.01\times 10^{-1}$ & $1.55\times 10^{2}$ \\
$3.16\times 10^7$ & $1.00\times10^{-3}$ & $3.16\times10^3$ & $5.62\times10^0$ & $3.14$ & $288^2\times144$  & 
$3.03\times10^{-2}$ & $2.30\times 10^{-1}$ & $1.45\times 10^{2}$ \\
$3.16\times 10^7$ & $3.16\times10^{-4}$ & $6.81\times10^2$ & $1.78\times10^0$ & $3.14$ & $288^2\times288$  & 
$3.10\times10^{-2}$ & $2.28\times 10^{-1}$ & $1.16\times 10^{2}$ \\
$3.16\times 10^7$ & $1.78\times10^{-4}$ & $3.16\times10^2$ & $1.00\times10^{0}$ & $3.14$ & $288^2\times288$  & 
$3.58\times10^{-2}$ & $1.95\times 10^{-1}$ & $9.42\times 10^{1}$ \\
$3.16\times 10^7$ & $1.00\times10^{-4}$ & $1.47\times10^2$ & $5.62\times10^{-1}$ & $3.14$ & $288^2\times288$  & 
$4.87\times10^{-2}$ & $1.23\times 10^{-1}$ & $6.14\times 10^{1}$ \\
$3.16\times 10^7$ & $5.62\times10^{-5}$ & $6.80\times10^1$ & $3.16\times10^{-1}$ & $3.14$ & $288^2\times288$  & 
$7.46\times10^{-2}$ & $2.71\times 10^{-2}$ & $2.28\times 10^{1}$ \\
$3.16\times 10^7$ & $3.16\times10^{-5}$ & $3.16\times10^1$ & $1.78\times10^{-1}$ & $3.14$ & $288^2\times288$  & 
\multicolumn{3}{c|}{Fully conductive} \\
\hline
$1.00\times 10^8$ & $\infty$ & - & - & $3.14$ & $288^2\times144$ & 
$2.56\times10^{-2}$ & $2.10\times 10^{-1}$ & $2.43\times 10^{2}$ \\
$1.00\times 10^8$ & $3.16\times10^{-3}$ & $4.65\times10^4$ & $3.16\times10^1$ & $3.14$ & $288^2\times144$  & 
$2.57\times10^{-2}$ & $2.15\times 10^{-1}$ & $2.44\times 10^{2}$ \\
$1.00\times 10^8$ & $1.78\times10^{-3}$ & $2.16\times10^4$ & $1.78\times10^1$ & $3.14$ & $288^2\times144$  & 
$2.56\times10^{-2}$ & $2.29\times 10^{-1}$ & $2.50\times 10^{2}$ \\
$1.00\times 10^8$ & $1.00\times10^{-3}$ & $1.00\times10^4$ & $1.00\times10^1$ & $3.14$ & $288^2\times144$  & 
$2.52\times10^{-2}$ & $2.46\times 10^{-1}$ & $2.44\times 10^{2}$ \\
$1.00\times 10^8$ & $5.62\times10^{-4}$ & $4.64\times10^3$ & $5.62\times10^0$ & $3.14$ & $288^2\times144$  & 
$2.42\times10^{-2}$ & $2.53\times 10^{-1}$ & $2.26\times 10^{2}$ \\
$1.00\times 10^8$ & $3.16\times10^{-4}$ & $2.16\times10^3$ & $3.16\times10^0$ & $3.14$ & $288^2\times288$  & 
$2.40\times10^{-2}$ & $2.57\times 10^{-1}$ & $2.06\times 10^{2}$ \\
$1.00\times 10^8$ & $1.78\times10^{-4}$ & $1.00\times10^3$ & $1.78\times10^0$ & $3.14$ & $288^2\times288$  & 
$2.45\times10^{-2}$ & $2.57\times 10^{-1}$ & $1.84\times 10^{2}$ \\
$1.00\times 10^8$ & $1.00\times10^{-4}$ & $4.64\times10^2$ & $1.00\times10^0$ & $3.14$ & $288^2\times288$  & 
$2.76\times10^{-2}$ & $2.34\times 10^{-1}$ & $1.56\times 10^{2}$ \\
$1.00\times 10^8$ & $5.62\times10^{-5}$ & $2.15\times10^2$ & $5.62\times10^{-1}$ & $3.14$ & $288^2\times288$  & 
$3.83\times10^{-2}$ & $1.68\times 10^{-1}$ & $1.08\times 10^{2}$ \\
$1.00\times 10^8$ & $3.16\times10^{-5}$ & $1.00\times10^2$ & $3.16\times10^{-1}$ & $3.14$ & $288^2\times288$  & 
$5.85\times10^{-2}$ & $1.15\times 10^{-1}$ & $6.11\times 10^{1}$ \\
$1.00\times 10^8$ & $1.78\times10^{-5}$ & $4.65\times10^1$ & $1.78\times10^{-1}$ & $3.14$ & $288^2\times288$  & 
\multicolumn{3}{c|}{Fully conductive} \\
\hline
$3.16\times 10^8$ & $\infty$ & - & - & $1.5$ & $288^2\times144$ & 
$2.06\times10^{-2}$ & $2.31\times 10^{-1}$ & $3.83\times 10^{2}$ \\
$3.16\times 10^8$ & $3.16\times10^{-3}$ & $1.47\times10^5$ & $5.63\times10^1$ & $1.5$ & $288^2\times144$  & 
$2.05\times10^{-2}$ & $2.33\times 10^{-1}$ & $3.89\times 10^{2}$ \\
$3.16\times 10^8$ & $1.00\times10^{-3}$ & $3.16\times10^4$ & $1.78\times10^1$ & $1.5$ & $288^2\times144$  & 
$2.06\times10^{-2}$ & $2.61\times 10^{-1}$ & $4.01\times 10^{2}$ \\
$3.16\times 10^8$ & $3.16\times10^{-4}$ & $6.81\times10^3$ & $5.63\times10^0$ & $1.5$ & $288^2\times288$  & 
$1.95\times10^{-2}$ & $2.75\times 10^{-1}$ & $3.46\times 10^{2}$ \\
$3.16\times 10^8$ & $1.00\times10^{-4}$ & $1.47\times10^3$ & $1.78\times10^{0}$ & $1.5$ & $288^2\times288$  & 
$1.96\times10^{-2}$ & $2.85\times 10^{-1}$ & $2.91\times 10^{2}$ \\
$3.16\times 10^8$ & $3.16\times10^{-5}$ & $3.16\times10^2$ & $5.62\times10^{-1}$ & $1.5$ & $288^2\times288$  & 
$3.02\times10^{-2}$ & $2.05\times 10^{-1}$ & $1.81\times 10^{2}$ \\
$3.16\times 10^8$ & $1.78\times10^{-5}$ & $1.47\times10^2$ & $3.16\times10^{-1}$ & $1.5$ & $288^2\times288$  & 
$4.60\times10^{-2}$ & $1.31\times 10^{-1}$ & $1.14\times 10^{2}$ \\
$3.16\times 10^8$ & $1.00\times10^{-5}$ & $6.81\times10^1$ & $1.78\times10^{-1}$ & $1.5$ & $288^2\times384$  & 
$7.56\times10^{-2}$ & $2.35\times 10^{-2}$ & $3.77\times 10^{1}$ \\
$3.16\times 10^8$ & $5.62\times10^{-6}$ & $3.16\times10^1$ & $1.00\times10^{-2}$ & $1.5$ & $288^2\times384$  & 
\multicolumn{3}{c|}{Fully conductive} \\
\hline 
$1.00\times 10^9$ & $\infty$ & - & - & $1.5$ & $288^2\times144$ & 
$1.64\times10^{-2}$ & $2.52\times 10^{-1}$ & $6.12\times 10^{2}$ \\
$1.00\times 10^9$ & $3.16\times10^{-3}$ & $4.65\times10^5$ & $1.00\times10^2$ & $1.5$ & $288^2\times144$  & 
$1.64\times10^{-2}$ & $2.54\times 10^{-1}$ & $6.06\times 10^{2}$ \\
$1.00\times 10^9$ & $1.78\times10^{-3}$ & $2.16\times10^5$ & $5.63\times10^1$ & $1.5$ & $288^2\times144$  & 
$1.64\times10^{-2}$ & $2.59\times 10^{-1}$ & $6.16\times 10^{2}$ \\
$1.00\times 10^9$ & $1.00\times10^{-3}$ & $1.00\times10^5$ & $3.16\times10^1$ & $1.5$ & $288^2\times144$  & 
$1.64\times10^{-2}$ & $2.69\times 10^{-1}$ & $6.19\times 10^{2}$ \\
$1.00\times 10^9$ & $5.62\times10^{-4}$ & $4.64\times10^4$ & $1.78\times10^1$ & $1.5$ & $288^2\times144$  & 
$1.61\times10^{-2}$ & $2.84\times 10^{-1}$ & $6.10\times 10^{2}$ \\
$1.00\times 10^9$ & $3.16\times10^{-4}$ & $2.16\times10^4$ & $1.00\times10^1$ & $1.5$ & $288^2\times288$  & 
$1.58\times10^{-2}$ & $2.90\times 10^{-1}$ & $5.68\times 10^{2}$ \\
$1.00\times 10^8$ & $1.78\times10^{-4}$ & $1.00\times10^4$ & $5.63\times10^0$ & $1.5$ & $288^2\times288$  & 
$1.56\times10^{-2}$ & $2.97\times 10^{-1}$ & $5.40\times 10^{2}$ \\
$1.00\times 10^9$ & $1.00\times10^{-4}$ & $4.64\times10^3$ & $3.16\times10^0$ & $1.5$ & $288^2\times288$  & 
$1.55\times10^{-2}$ & $2.99\times 10^{-1}$ & $4.81\times 10^{2}$ \\
$1.00\times 10^9$ & $5.62\times10^{-5}$ & $2.15\times10^3$ & $1.78\times10^0$ & $1.5$ & $288^2\times384$  & 
$1.57\times10^{-2}$ & $2.96\times 10^{-1}$ & $4.40\times 10^{2}$ \\
$1.00\times 10^9$ & $3.16\times10^{-5}$ & $1.00\times10^3$ & $1.00\times10^0$ & $1.5$ & $288^2\times384$  & 
$1.74\times10^{-2}$ & $2.89\times 10^{-1}$ & $3.86\times 10^{2}$ \\
$1.00\times 10^9$ & $1.78\times10^{-5}$ & $4.65\times10^2$ & $5.63\times10^{-1}$ & $1.5$ & $288^2\times384$  & 
$2.34\times10^{-2}$ & $2.38\times 10^{-1}$ & $2.95\times 10^{2}$ \\
$1.00\times 10^9$ & $1.00\times10^{-5}$ & $2.15\times10^2$ & $3.16\times10^{-1}$ & $1.5$ & $288^2\times512$  & 
$3.58\times10^{-2}$ & $1.73\times 10^{-1}$ & $2.00\times 10^{2}$ \\
$1.00\times 10^9$ & $5.62\times10^{-6}$ & $1.00\times10^2$ & $1.78\times10^{-1}$ & $1.5$ & $288^2\times512$  & 
$5.96\times10^{-2}$ & $8.79\times 10^{-2}$ & $1.10\times 10^{2}$ \\
$1.00\times 10^9$ & $3.16\times10^{-6}$ & $4.65\times10^1$ & $1.00\times10^{-1}$ & $1.5$ & $288^2\times512$  & 
\multicolumn{3}{c|}{Fully conductive} \\
\hline
$3.16\times 10^9$ & $\infty$ & - & - &  $1$ & $384^2\times288$ & 
$1.30\times10^{-2}$ & $2.69\times 10^{-1}$ & $9.16\times 10^{2}$ \\
$3.16\times 10^9$ & $3.16\times10^{-3}$ & $1.47\times10^6$ & $1.78\times10^2$ & $1$ & $384^2\times288$  & 
$1.29\times10^{-2}$ & $2.72\times 10^{-1}$ & $9.16\times 10^{2}$ \\
$3.16\times 10^9$ & $1.00\times10^{-3}$ & $3.16\times10^5$ & $5.62\times10^1$ & $1$ & $384^2\times288$  & 
$1.29\times10^{-2}$ & $2.72\times 10^{-1}$ & $9.18\times 10^{2}$ \\
$3.16\times 10^9$ & $3.16\times10^{-4}$ & $6.81\times10^4$ & $1.78\times10^1$ & $1$ & $384^2\times288$  & 
$1.28\times10^{-2}$ & $2.98\times 10^{-1}$ & $9.19\times 10^{2}$ \\
$3.16\times 10^9$ & $1.00\times10^{-4}$ & $1.47\times10^4$ & $5.62\times10^0$ & $1$ & $384^2\times288$  & 
$1.23\times10^{-2}$ & $3.16\times 10^{-1}$ & $7.94\times 10^{2}$ \\
$3.16\times 10^9$ & $3.16\times10^{-5}$ & $3.16\times10^3$ & $1.78\times10^{0}$ & $1$ & $384^2\times384$  & 
$1.29\times10^{-2}$ & $3.18\times 10^{-1}$ & $6.63\times 10^{2}$ \\
$3.16\times 10^9$ & $1.00\times10^{-5}$ & $6.81\times10^2$ & $5.62\times10^{-1}$ & $1$ & $384^2\times512$  & 
$1.84\times10^{-2}$ & $2.71\times 10^{-1}$ & $4.78\times 10^{2}$ \\
$3.16\times 10^9$ & $5.62\times10^{-6}$ & $3.16\times10^2$ & $3.16\times10^{-1}$ & $1$ & $384^2\times512$  & 
$2.81\times10^{-2}$ & $2.07\times 10^{-1}$ & $3.32\times 10^{2}$ \\
$3.16\times 10^9$ & $3.16\times10^{-6}$ & $1.47\times10^2$ & $1.78\times10^{-1}$ & $1$ & $384^2\times512$  & 
$4.62\times10^{-2}$ & $1.25\times 10^{-1}$ & $1.98\times 10^{2}$ \\
$3.16\times 10^9$ & $1.78\times10^{-6}$ & $6.83\times10^1$ & $1.00\times10^{-1}$ & $1$ & $384^2\times512$  & 
$7.77\times10^{-2}$ & $2.05\times 10^{-2}$ & $6.31\times 10^{1}$ \\
$3.16\times 10^9$ & $1.00\times10^{-6}$ & $3.16\times10^1$ & $5.62\times10^{-2}$ & $1$ & $384^2\times512$  & 
\multicolumn{3}{c|}{Fully conductive} \\
\hline 
$1.00\times 10^{10}$ & $\infty$ & - & - & $1.5$ & $384^2\times288$ & 
$1.01\times10^{-2}$ & $2.77\times 10^{-1}$ & $1.41\times 10^{3}$ \\
$1.00\times 10^{10}$ & $3.16\times10^{-3}$ & $4.65\times10^6$ & $3.16\times10^2$ & $1$ & $384^2\times288$  & 
$1.00\times10^{-3}$ & $2.83\times 10^{-1}$ & $1.42\times 10^{3}$ \\
$1.00\times 10^{10}$ & $1.78\times10^{-3}$ & $2.16\times10^6$ & $1.78\times10^2$ & $1$ & $384^2\times288$  & 
$1.00\times10^{-2}$ & $2.83\times 10^{-1}$ & $1.42\times 10^{3}$ \\
$1.00\times 10^{10}$ & $1.00\times10^{-3}$ & $1.00\times10^6$ & $1.00\times10^2$ & $1$ & $384^2\times288$  & 
$1.01\times10^{-2}$ & $2.80\times 10^{-1}$ & $1.40\times 10^{3}$ \\
$1.00\times 10^{10}$ & $5.62\times10^{-4}$ & $4.64\times10^5$ & $5.62\times10^1$ & $1$ & $384^2\times288$  & 
$1.01\times10^{-2}$ & $2.90\times 10^{-1}$ & $1.46\times 10^{3}$ \\
$1.00\times 10^{10}$ & $3.16\times10^{-4}$ & $2.16\times10^5$ & $3.16\times10^1$ & $1$ & $384^2\times288$  & 
$1.00\times10^{-2}$ & $3.17\times 10^{-1}$ & $1.45\times 10^{3}$ \\
$1.00\times 10^{10}$ & $1.78\times10^{-4}$ & $1.00\times10^5$ & $1.78\times10^1$ & $1$ & $384^2\times288$  & 
$1.00\times10^{-2}$ & $3.33\times 10^{-1}$ & $1.41\times 10^{3}$ \\
$1.00\times 10^{10}$ & $1.00\times10^{-4}$ & $4.64\times10^4$ & $1.00\times10^1$ & $1$ & $384^2\times384$  & 
$9.84\times10^{-3}$ & $3.36\times 10^{-1}$ & $1.29\times 10^{3}$ \\
$1.00\times 10^{10}$ & $5.62\times10^{-5}$ & $2.15\times10^4$ & $5.62\times10^0$ & $1$ & $384^2\times384$  & 
$9.82\times10^{-3}$ & $3.38\times 10^{-1}$ & $1.21\times 10^{3}$ \\
$1.00\times 10^{10}$ & $3.16\times10^{-5}$ & $1.00\times10^4$ & $3.16\times10^0$ & $1$ & $384^2\times384$  & 
$1.01\times10^{-2}$ & $3.35\times 10^{-1}$ & $1.11\times 10^{3}$ \\
$1.00\times 10^{10}$ & $1.78\times10^{-5}$ & $4.65\times10^3$ & $1.78\times10^0$ & $1$ & $384^2\times384$  & 
$1.05\times10^{-2}$ & $3.28\times 10^{-1}$ & $9.67\times 10^{2}$ \\
$1.00\times 10^{10}$ & $1.00\times10^{-5}$ & $2.15\times10^3$ & $1.00\times10^0$ & $1$ & $384^2\times512$  & 
$1.07\times10^{-2}$ & $3.18\times 10^{-1}$ & $8.82\times 10^{2}$ \\
$1.00\times 10^{10}$ & $5.62\times10^{-6}$ & $1.00\times10^3$ & $5.62\times10^{-1}$ & $1$ & $384^2\times512$  & 
$1.44\times10^{-2}$ & $2.93\times 10^{-1}$ & $7.40\times 10^{2}$ \\
$1.00\times 10^{10}$ & $3.16\times10^{-6}$ & $4.65\times10^2$ & $3.16\times10^{-1}$ & $1$ & $384^2\times512$  & 
$2.18\times10^{-2}$ & $2.41\times 10^{-1}$ & $5.47\times 10^{3}$ \\
$1.00\times 10^{10}$ & $1.78\times10^{-6}$ & $2.16\times10^2$ & $1.78\times10^{-1}$ & $1$ & $384^2\times512$  & 
$3.61\times10^{-2}$ & $1.71\times 10^{-1}$ & $3.59\times 10^{3}$ \\
$1.00\times 10^{10}$ & $1.00\times10^{-6}$ & $1.00\times10^2$ & $1.00\times10^{-1}$ & $1$ & $384^2\times512$  & 
$6.11\times10^{-2}$ & $7.78\times 10^{-1}$ & $1.82\times 10^{3}$ \\
\hline
\end{longtable}
\end{landscape}

\bibliographystyle{jfm}
\bibliography{jfm}

\begin{thebibliography}{53}
\expandafter\ifx\csname natexlab\endcsname\relax\def\natexlab#1{#1}\fi
\def\au#1{#1} \def\ed#1{#1} \def\yr#1{#1}\def\at#1{#1}\def\jt#1{\textit{#1}} \def\bt#1{#1}\def\bvol#1{\textbf{#1}} \def\vol#1{#1} \def\pg#1{#1} \def\publ#1{#1}\def\arxiv#1{#1}\def\org#1{#1}\def\st#1{\textit{#1}}

\bibitem[Aguirre~Guzm{\'a}n {\em et~al.\/}(2022)Aguirre~Guzm{\'a}n, Madonia, Cheng, Ostilla-M{\'o}nico, Clercx \& Kunnen]{aguirre2022flow}
{\sc \au{Aguirre~Guzm{\'a}n, A.~J.}, \au{Madonia, M.}, \au{Cheng, J.~S.}, \au{Ostilla-M{\'o}nico, R.}, \au{Clercx, H. J.~H.} \& \au{Kunnen, R. P.~J.}} \yr{2022}  \at{Flow-and temperature-based statistics characterizing the regimes in rapidly rotating turbulent convection in simulations employing no-slip boundary conditions}.  \jt{Physical Review Fluids}  \bvol{7}~(1),  \pg{013501}.

\bibitem[Arslan(2024)]{arslan2024internally}
{\sc \au{Arslan, A.}} \yr{2024}  \at{Internally heated convection with rotation: bounds on heat transport}.  \jt{Journal of Fluid Mechanics}  \bvol{1001},  \pg{A55}.

\bibitem[Arslan {\em et~al.\/}(2021{\natexlab{{\em a\/}}})Arslan, Fantuzzi, Craske \& Wynn]{Arslan2021a}
{\sc \au{Arslan, A.}, \au{Fantuzzi, G.}, \au{Craske, J.} \& \au{Wynn, A.}} \yr{2021{\natexlab{{\em a\/}}}}  \at{{Bounds for internally heated convection with fixed boundary heat flux}}.  \jt{Journal of Fluid Mechanics}  \bvol{992},  \pg{R1}.

\bibitem[Arslan {\em et~al.\/}(2021{\natexlab{{\em b\/}}})Arslan, Fantuzzi, Craske \& Wynn]{Arslan2021}
{\sc \au{Arslan, A.}, \au{Fantuzzi, G.}, \au{Craske, J.} \& \au{Wynn, A.}} \yr{2021{\natexlab{{\em b\/}}}}  \at{{Bounds on heat transport for convection driven by internal heating}}.  \jt{Journal of Fluid Mechanics}  \bvol{919},  \pg{A15}.

\bibitem[Arslan {\em et~al.\/}(2024)Arslan, Fantuzzi, Craske \& Wynn]{Arslan2024b}
{\sc \au{Arslan, A.}, \au{Fantuzzi, G.}, \au{Craske, J.} \& \au{Wynn, A.}} \yr{2024}  \at{Internal heating profiles for which downward conduction is impossible}.  \jt{Journal of Fluid Mechanics}  \bvol{993},  \pg{A5}.

\bibitem[Arslan \& Rojas(2025)]{arslan2025a}
{\sc \au{Arslan, Ali} \& \au{Rojas, Rubén~E.}} \yr{2025}  \at{New bounds for heat transport in internally heated convection at infinite prandtl number}.  \jt{Journal of Mathematical Physics}  \bvol{66}~(2),  \pg{023101}.

\bibitem[Bouillaut {\em et~al.\/}(2022)Bouillaut, Flesselles, Miquel, Aumaître \& Gallet]{bouillaut2022}
{\sc \au{Bouillaut, V.}, \au{Flesselles, B.}, \au{Miquel, B.}, \au{Aumaître, S.} \& \au{Gallet, B.}} \yr{2022}  \at{Velocity-informed upper bounds on the convective heat transport induced by internal heat sources and sinks}.  \jt{Philosophical Transactions of the Royal Society A}  \bvol{380}~(2225),  \pg{20210034}.

\bibitem[Bouillaut {\em et~al.\/}(2019)Bouillaut, Lepot, Auma{\^{i}}tre \& Gallet]{Bouillaut2019}
{\sc \au{Bouillaut, V.}, \au{Lepot, S.}, \au{Auma{\^{i}}tre, S.} \& \au{Gallet, B.}} \yr{2019}  \at{{Transition to the ultimate regime in a radiatively driven convection experiment}}.  \jt{Journal of Fluid Mechanics}  \bvol{861}.

\bibitem[Chandrasekhar(1961)]{chandrasekhar1961hydrodynamic}
{\sc \au{Chandrasekhar, S.}} \yr{1961} {\em Hydrodynamic and hydromagnetic stability\/}.  \publ{Oxford University Pressn}.

\bibitem[Chernyshenko(2022)]{Chernyshenko2022}
{\sc \au{Chernyshenko, S.}} \yr{2022}  \at{{Relationship between the methods of bounding time averages}}.  \jt{Philosophical Transactions of the Royal Society A}  \bvol{380}~(1),  \pg{20210044.}

\bibitem[Constantin {\em et~al.\/}(1999)Constantin, Hallstrom \& Putkaradze]{constantin1999r}
{\sc \au{Constantin, P.}, \au{Hallstrom, C.} \& \au{Putkaradze, V.}} \yr{1999}  \at{Heat transport in rotating convection}.  \jt{Physica D: Nonlinear Phenomena}  \bvol{125}~(3-4),  \pg{275--284}.

\bibitem[Doering(2019)]{Doering2019}
{\sc \au{Doering, C.~R.}} \yr{2019}  \at{Thermal forcing and ‘classical’ and ‘ultimate’ regimes of rayleigh–bénard convection}.  \jt{Journal of Fluid Mechanics}  \bvol{868},  \pg{1–4}.

\bibitem[Doering \& Constantin(2001)]{Doering2001}
{\sc \au{Doering, C.~R.} \& \au{Constantin, P.}} \yr{2001}  \at{{On upper bounds for infinite Prandtl number convection with or without rotation}}.  \jt{Journal of Mathematical Physics}  \bvol{42}~(2),  \pg{784--795}.

\bibitem[Dormy(2025)]{dormy2025rapidly}
{\sc \au{Dormy, E.}} \yr{2025}  \at{Rapidly rotating magnetohydrodynamics and the geodynamo}.  \jt{Annual Review of Fluid Mechanics}  \bvol{57}.

\bibitem[Ecke \& Shishkina(2023)]{ecke2023}
{\sc \au{Ecke, R.~E.} \& \au{Shishkina, O.}} \yr{2023}  \at{Turbulent rotating rayleigh--b{\'e}nard convection}.  \jt{Annual review of fluid mechanics}  \bvol{55},  \pg{603--638}.

\bibitem[Fantuzzi {\em et~al.\/}(2022)Fantuzzi, Arslan \& Wynn]{Fantuzzi2022}
{\sc \au{Fantuzzi, G.}, \au{Arslan, A.} \& \au{Wynn, A.}} \yr{2022}  \at{{The background method: Theory and computations}}.  \jt{Philosophical Transactions of the Royal Society A}  \bvol{380}~(1),  \pg{20210038}.

\bibitem[Fantuzzi {\em et~al.\/}(2016)Fantuzzi, Goluskin, Huang \& Chernyshenko]{Fantuzzi2016b}
{\sc \au{Fantuzzi, G.}, \au{Goluskin, D.}, \au{Huang, D.} \& \au{Chernyshenko, S.~I.}} \yr{2016}  \at{{Bounds for deterministic and stochastic dynamical systems using sum-of-squares optimization}}.  \jt{SIAM Journal on Applied Dynamical Systems}  \bvol{15}~(4),  \pg{1962--1988}.

\bibitem[Gallet(2015)]{gallet2015}
{\sc \au{Gallet, B.}} \yr{2015}  \at{Exact two-dimensionalization of rapidly rotating large-reynolds-number flows}.  \jt{Journal of Fluid Mechanics}  \bvol{783},  \pg{412--447}.

\bibitem[Gastine {\em et~al.\/}(2016)Gastine, Wicht \& Aubert]{gastine2016scaling}
{\sc \au{Gastine, T.}, \au{Wicht, J.} \& \au{Aubert, J.}} \yr{2016}  \at{Scaling regimes in spherical shell rotating convection}.  \jt{Journal of Fluid Mechanics}  \bvol{808},  \pg{690--732}.

\bibitem[Goluskin(2016)]{Goluskin2016book}
{\sc \au{Goluskin, D.}} \yr{2016} {\em {Internally heated convection and Rayleigh--B\'enard convection}\/}. {\em Springer Briefs in Applied Sciences and Technologies\/} .  \publ{Springer, Cham}.

\bibitem[Goluskin \& Van~der Poel(2016)]{goluskin2016penetrative}
{\sc \au{Goluskin, D.} \& \au{Van~der Poel, E.~P.}} \yr{2016}  \at{Penetrative internally heated convection in two and three dimensions}.  \jt{Journal of fluid mechanics}  \bvol{791},  \pg{R6}.

\bibitem[Goluskin \& Spiegel(2012)]{goluskin2012convection}
{\sc \au{Goluskin, D.} \& \au{Spiegel, E.~A.}} \yr{2012}  \at{{Convection driven by internal heating}}.  \jt{Physics Letters A}  \bvol{377}~(1-2),  \pg{83--92}.

\bibitem[Goulart \& Chernyshenko(2012)]{goulart2012global}
{\sc \au{Goulart, P.~J.} \& \au{Chernyshenko, S.~I.}} \yr{2012}  \at{{Global stability analysis of fluid flows using sum-of-squares}}.  \jt{Physica D: Nonlinear Phenomena}  \bvol{241}~(6),  \pg{692--704}.

\bibitem[Greenspan(1968)]{greenspan1968}
{\sc \au{Greenspan, H.~P.}} \yr{1968} {\em The theory of rotating fluids\/}.  \publ{Cambridge University Press}.

\bibitem[Grooms \& Whitehead(2014)]{grooms2014}
{\sc \au{Grooms, I.} \& \au{Whitehead, J.~P.}} \yr{2014}  \at{Bounds on heat transport in rapidly rotating rayleigh--b{\'e}nard convection}.  \jt{Nonlinearity}  \bvol{28}~(1),  \pg{29}.

\bibitem[Grossmann \& Lohse(2000)]{grossmann2000scaling}
{\sc \au{Grossmann, S.} \& \au{Lohse, D.}} \yr{2000}  \at{Scaling in thermal convection: a unifying theory}.  \jt{Journal of Fluid Mechanics}  \bvol{407},  \pg{27--56}.

\bibitem[Guervilly \& Cardin(2016)]{guervilly2016}
{\sc \au{Guervilly, C.} \& \au{Cardin, P.}} \yr{2016}  \at{Subcritical convection of liquid metals in a rotating sphere using a quasi-geostrophic model}.  \jt{Journal of Fluid Mechanics}  \bvol{808},  \pg{61--89}.

\bibitem[Hadjerci {\em et~al.\/}(2024)Hadjerci, Bouillaut, Miquel \& Gallet]{hadjerci2024}
{\sc \au{Hadjerci, G.}, \au{Bouillaut, V.}, \au{Miquel, B.} \& \au{Gallet, B.}} \yr{2024}  \at{Rapidly rotating radiatively driven convection: experimental and numerical validation of the ‘geostrophic turbulence’scaling predictions}.  \jt{Journal of Fluid Mechanics}  \bvol{998},  \pg{A9}.

\bibitem[Jones {\em et~al.\/}(2000)Jones, Soward \& Mussa]{jones2000}
{\sc \au{Jones, C.~A.}, \au{Soward, A.~M.} \& \au{Mussa, A.~I.}} \yr{2000}  \at{The onset of thermal convection in a rapidly rotating sphere}.  \jt{Journal of Fluid Mechanics}  \bvol{405},  \pg{157--179}.

\bibitem[Kaplan {\em et~al.\/}(2017)Kaplan, Schaeffer, Vidal \& Cardin]{kaplan2017}
{\sc \au{Kaplan, E.~J.}, \au{Schaeffer, N.}, \au{Vidal, J.} \& \au{Cardin, P.}} \yr{2017}  \at{Subcritical thermal convection of liquid metals in a rapidly rotating sphere}.  \jt{Physical review letters}  \bvol{119}~(9),  \pg{094501}.

\bibitem[Kazemi {\em et~al.\/}(2022)Kazemi, Ostilla-M{\'o}nico \& Goluskin]{kazemi2022transition}
{\sc \au{Kazemi, S.}, \au{Ostilla-M{\'o}nico, R.} \& \au{Goluskin, D.}} \yr{2022}  \at{Transition between boundary-limited scaling and mixing-length scaling of turbulent transport in internally heated convection}.  \jt{Physical Review Letters}  \bvol{129}~(2),  \pg{024501}.

\bibitem[King \& Aurnou(2012)]{king2012thermal}
{\sc \au{King, E.~M.} \& \au{Aurnou, J.~M.}} \yr{2012}  \at{Thermal evidence for taylor columns in turbulent rotating rayleigh-b{\'e}nard convection}.  \jt{Physical Review E—Statistical, Nonlinear, and Soft Matter Physics}  \bvol{85}~(1),  \pg{016313}.

\bibitem[King {\em et~al.\/}(2009)King, Stellmach, Noir, Hansen \& Aurnou]{king2009boundary}
{\sc \au{King, E.~M.}, \au{Stellmach, S.}, \au{Noir, J.}, \au{Hansen, U.} \& \au{Aurnou, J.~M.}} \yr{2009}  \at{Boundary layer control of rotating convection systems}.  \jt{Nature}  \bvol{457}~(7227),  \pg{301--304}.

\bibitem[Kumar {\em et~al.\/}(2022)Kumar, Arslan, Fantuzzi, Craske \& Wynn]{kumar2021ihc}
{\sc \au{Kumar, A.}, \au{Arslan, A.}, \au{Fantuzzi, G.}, \au{Craske, J.} \& \au{Wynn, A.}} \yr{2022}  \at{Analytical bounds on the heat transport in internally heated convection}.  \jt{Journal of Fluid Mechanics}  \bvol{938},  \pg{A26}.

\bibitem[Kunnen(2021)]{Kunnen04052021}
{\sc \au{Kunnen, Rudie P.~J.}} \yr{2021}  \at{The geostrophic regime of rapidly rotating turbulent convection}.  \jt{Journal of Turbulence}  \bvol{22}~(4-5),  \pg{267--296},  \arxiv{arXiv: https://doi.org/10.1080/14685248.2021.1876877}.

\bibitem[Kunnen {\em et~al.\/}(2016)Kunnen, Ostilla-M{\'o}nico, Van Der~Poel, Verzicco \& Lohse]{kunnen2016transition}
{\sc \au{Kunnen, R. P.~J.}, \au{Ostilla-M{\'o}nico, R.}, \au{Van Der~Poel, E.~P.}, \au{Verzicco, R.} \& \au{Lohse, D.}} \yr{2016}  \at{Transition to geostrophic convection: the role of the boundary conditions}.  \jt{Journal of Fluid Mechanics}  \bvol{799},  \pg{413--432}.

\bibitem[Lepot {\em et~al.\/}(2018)Lepot, Auma{\^{i}}tre \& Gallet]{Lepot2018}
{\sc \au{Lepot, S.}, \au{Auma{\^{i}}tre, S.} \& \au{Gallet, B.}} \yr{2018}  \at{{Radiative heating achieves the ultimate regime of thermal convection}}.  \jt{Proceedings of the National Academy of Sciences of the U.S.A.}  \bvol{115}~(36),  \pg{8937--8941}.

\bibitem[Lohse \& Shishkina(2023)]{lohse2023}
{\sc \au{Lohse, D.} \& \au{Shishkina, O.}} \yr{2023}  \at{Ultimate turbulent thermal convection}.  \jt{Physics Today 1}  \bvol{76(11)},  \pg{26--32}.

\bibitem[Miquel {\em et~al.\/}(2019)Miquel, Lepot, Bouillaut \& Gallet]{Miquel2019}
{\sc \au{Miquel, B.}, \au{Lepot, S.}, \au{Bouillaut, V.} \& \au{Gallet, B.}} \yr{2019}  \at{{Convection driven by internal heat sources and sinks: Heat transport beyond the mixing-length or "ultimate" scaling regime}}.  \jt{Physical Review Fluids}  \bvol{4}~(12),  \pg{121501}.

\bibitem[Pachev {\em et~al.\/}(2020)Pachev, Whitehead, Fantuzzi \& Grooms]{Pachev2020}
{\sc \au{Pachev, B.}, \au{Whitehead, J.~P.}, \au{Fantuzzi, G.} \& \au{Grooms, I.}} \yr{2020}  \at{{Rigorous bounds on the heat transport of rotating convection with Ekman pumping}}.  \jt{Journal of Mathematical Physics}  \bvol{61}~(2),  \pg{023101}.

\bibitem[Pedlosky(2013)]{pedlosky}
{\sc \au{Pedlosky, J.}} \yr{2013} {\em Geophysical fluid dynamics\/}.  \publ{Springer Science \& Business Media}.

\bibitem[Ricard(2015)]{ricard2015}
{\sc \au{Ricard, Y.}} \yr{2015}  \at{7.02 - physics of mantle convection}. In {\em Treatise on Geophysics (Second Edition)\/},  \bt{Second edition edn. (ed. \ed{Gerald Schubert})},  \pg{pp. 23--71}.  \publ{Oxford: Elsevier}.

\bibitem[Roberts(2015)]{ROBERTS201557}
{\sc \au{Roberts, P.H.}} \yr{2015}  \at{8.03 - theory of the geodynamo}. In {\em Treatise on Geophysics (Second Edition)\/},  \bt{Second edition edn. (ed. \ed{Gerald Schubert})},  \pg{pp. 57--90}.  \publ{Oxford: Elsevier}.

\bibitem[Roberts(1967)]{roberts1967convection}
{\sc \au{Roberts, P.~H.}} \yr{1967}  \at{Convection in horizontal layers with internal heat generation. {{{Theory}}}}.  \jt{Journal of Fluid Mechanics}  \bvol{30}~(1),  \pg{33--49}.

\bibitem[Rosa \& Temam(2022)]{Rosa2020}
{\sc \au{Rosa, R. M.~S.} \& \au{Temam, R.~M.}} \yr{2022}  \at{{Optimal minimax bounds for time and ensemble averages of dissipative infinite-dimensional systems with applications to the incompressible Navier--Stokes equations}}.  \jt{Pure and Applied Functional Analysis}  \bvol{7}~(1),  \pg{327--355}.

\bibitem[Schubert {\em et~al.\/}(2001)Schubert, Turcotte \& Olson]{schubert2001mantle}
{\sc \au{Schubert, G.}, \au{Turcotte, D.~L.} \& \au{Olson, P.}} \yr{2001} {\em Mantle convection in the Earth and planets\/}.  \publ{Cambridge University Press}.

\bibitem[Song {\em et~al.\/}(2024)Song, Shishkina \& Zhu]{Song2024}
{\sc \au{Song, J.}, \au{Shishkina, O.} \& \au{Zhu, X.}} \yr{2024}  \at{Scaling regimes in rapidly rotating thermal convection at extreme rayleigh numbers}.  \jt{Journal of Fluid Mechanics}  \bvol{984},  \pg{A45}.

\bibitem[Tilgner(2015)]{TILGNER2015183}
{\sc \au{Tilgner, A.}} \yr{2015}  \at{8.07 - rotational dynamics of the core}. In {\em Treatise on Geophysics (Second Edition)\/},  \bt{Second edition edn. (ed. \ed{Gerald Schubert})},  \pg{pp. 183--212}.  \publ{Oxford: Elsevier}.

\bibitem[Tilgner(2022)]{tilgner2022}
{\sc \au{Tilgner, A.}} \yr{2022}  \at{Bounds for rotating rayleigh--b{\'e}nard convection at large prandtl number}.  \jt{Journal of Fluid Mechanics}  \bvol{930},  \pg{A33}.

\bibitem[Tobasco {\em et~al.\/}(2018)Tobasco, Goluskin \& Doering]{Tobasco2018}
{\sc \au{Tobasco, I.}, \au{Goluskin, D.} \& \au{Doering, C.~R.}} \yr{2018}  \at{{Optimal bounds and extremal trajectories for time averages in nonlinear dynamical systems}}.  \jt{Physics Letters A}  \bvol{382}~(6),  \pg{382--386},  \arxiv{arXiv: 1705.07096}.

\bibitem[Van Der~Poel {\em et~al.\/}(2015)Van Der~Poel, Ostilla-M{\'o}nico, Donners \& Verzicco]{van2015pencil}
{\sc \au{Van Der~Poel, E.~P.}, \au{Ostilla-M{\'o}nico, R.}, \au{Donners, J.} \& \au{Verzicco, R.}} \yr{2015}  \at{A pencil distributed finite difference code for strongly turbulent wall-bounded flows}.  \jt{Computers \& Fluids}  \bvol{116},  \pg{10--16}.

\bibitem[Wang {\em et~al.\/}(2020)Wang, Lohse \& Shishkina]{Wang2020}
{\sc \au{Wang, Q.}, \au{Lohse, D.} \& \au{Shishkina, O.}} \yr{2020}  \at{Scaling in internally heated convection: a unifying theory}.  \jt{Geophysical Research Letters}  \bvol{47},  \pg{e2020GL091198}.

\bibitem[Yan(2004)]{Yan2004}
{\sc \au{Yan, X.}} \yr{2004}  \at{{On limits to convective heat transport at infinite Prandtl number with or without rotation}}.  \jt{Journal of Mathematical Physics}  \bvol{45}~(7),  \pg{2718--2743}.

\end{thebibliography}

\end{document}